\documentclass[sigconf,10pt]{acmart}
\acmSubmissionID{657}
\usepackage{subcaption}
\usepackage{xspace}
\usepackage{titlesec}

\titlespacing*{\section}{2pt}{2pt}{2pt}
\titlespacing*{\subsection}{2pt}{2pt}{2pt}

\usepackage{amsmath}
\usepackage[subtle]{savetrees}

\usepackage{algorithm}
\usepackage{algorithmic}

\usepackage{multirow}
\usepackage{comment}
\usepackage{fancybox, fancyvrb, calc}
\usepackage[inline]{enumitem}
\usepackage{url}
\usepackage[normalem]{ulem}
\usepackage{pbox}
\usepackage{wrapfig}
\usepackage[T1]{fontenc}
\usepackage[utf8]{inputenc}
\usepackage{mathtools}
\usepackage{enumitem}
\setlist{nolistsep}
\usepackage{hhline}
\usepackage{appendix}
\usepackage{outlines}
\usepackage[english]{babel}
\usepackage{hyperref}
\hypersetup{
  colorlinks=true,      
  linkcolor=blue,       
  citecolor=magenta,    
  filecolor=cyan,       
  urlcolor=blue          
}
\usepackage[hyperref]{backref}
\makeatletter
\def\maxwidth{ %
  \ifdim\Gin@nat@width>\linewidth
    \linewidth
  \else
    \Gin@nat@width
  \fi
}
\makeatother

\definecolor{fgcolor}{rgb}{0.345, 0.345, 0.345}

\usepackage{framed}
\makeatletter
 {\par\unskip\endMakeFramed%
 \at@end@of@kframe}
\makeatother

\definecolor{shadecolor}{rgb}{.97, .97, .97}
\definecolor{messagecolor}{rgb}{0, 0, 0}
\definecolor{warningcolor}{rgb}{1, 0, 1}
\definecolor{errorcolor}{rgb}{1, 0, 0}
\newenvironment{knitrout}{}{} 
\usepackage{alltt}
\usepackage{listings}
\newcommand{\cut}[1]{}
\newcommand{\paragrapha}[1]{\vspace{0.05in}\noindent{\bf #1.}}
\newcommand{\paragraphi}[1]{\noindent\textit{#1.}}

\newcommand{\Para}[1]{\paragrapha{#1}}
\newcommand{\Sec}[1]{\S\ref{#1}}
\newcommand{\eg}{e.g., }

\newcommand{\ie}{i.e., }
\newcommand{\etal}{\emph{et al.}\xspace}

\newcommand{\an}[1]{{\color{teal}\bf AN: {#1}}}

\newcommand{\fc}[1]{{\color{blue}\bf FC: {#1}}}
\newcommand{\radhika}[1]{{\color{magenta}\bf [RM: {#1}]}}
\newcommand{\bundle}{bundle\xspace}
\newcommand{\name}{Bundler\xspace}
\newcommand{\inbox}{sendbox\xspace}
\newcommand{\outbox}{receivebox\xspace}
\newcommand{\capinbox}{Sendbox\xspace}
\newcommand{\capoutbox}{Receivebox\xspace}
\newcommand{\pair}{\inbox-\outbox pair\xspace}

\usepackage[absolute]{textpos}

\titleformat{\section}
  {\normalfont\Large\bfseries}{\thesection}{1em}{\MakeUppercase}

\setcopyright{rightsretained}

\begin{document}
\fancyhead{}

\def\forArxiv{1}
\ifdefined\forArxiv
\begin{textblock}{13.5}(1,0.25)\fbox{
\begin{minipage}{\dimexpr\textwidth-2\fboxsep-2\fboxrule\relax}
        \scriptsize
        If you cite this paper, please use the EuroSys reference:
        Frank Cangialosi, Akshay Narayan, Prateesh Goyal, Radhika Mittal, Mohammad Alizadeh, and Hari Balakrishnan. 2021. Site-to-Site Internet Traffic Control. In \emph{EuroSys '21, April 26--28, 2021, Online, United Kingdom}. ACM, New York, NY,  USA, 16 pages. https://doi.org/10.1145/3447786.3456260
\end{minipage}}\end{textblock}
\fi

\copyrightyear{2021}
\acmYear{2021}
\acmConference[EuroSys '21]{Sixteenth European Conference on Computer Systems}{April 26--28, 2021}{Online, United Kingdom}
\acmBooktitle{Sixteenth European Conference on Computer Systems (EuroSys '21), April 26--28, 2021, Online, United Kingdom}\acmDOI{10.1145/3447786.3456260}
\acmISBN{978-1-4503-8334-9/21/04}

\renewcommand*{\thefootnote}{\fnsymbol{footnote}}

\newcommand{\beaver}{\includegraphics[width=15pt]{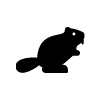}}
\newcommand{\corn}{\includegraphics[width=15pt]{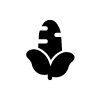}}

\title{\vspace{-10pt}\Huge Site-to-Site Internet Traffic Control}
\author{
    \vspace{-20pt}
    \Large
Frank Cangialosi\footnotemark[1]\beaver,
Akshay Narayan\footnotemark[1]\beaver,
Prateesh Goyal\beaver, \\\vspace{-5pt}
Radhika Mittal\corn,
Mohammad Alizadeh\beaver,
Hari Balakrishnan\beaver
}
\affiliation{\beaver MIT CSAIL \corn UIUC}

\renewcommand{\shortauthors}{Cangialosi and Narayan, et al.}

\date{\vspace{-12mm}}
\begin{abstract}
Queues allow network operators to control traffic: where queues build, they can enforce scheduling and shaping policies.
In the Internet today, however, there is a mismatch between where queues build and where control is most effectively enforced; queues build at bottleneck links that are often not under the control of the data sender. To resolve this mismatch, we propose a new kind of middlebox, called \name.
\name uses a novel inner control loop between a {\em sendbox} (in the sender's site) and a {\em receivebox} (in the receiver's site) to determine the aggregate rate for the bundle, leaving the end-to-end connections and their control loops intact. Enforcing this sending rate ensures that bottleneck queues that would have built up from the bundle's packets now shift from the bottleneck to the {\em \inbox}. 
This enables the sendbox to exercise control over its traffic by scheduling packets according to any policy necessary to achieve the network operator's higher-level objectives.
We have implemented \name in Linux and evaluated it with real-world and emulation experiments.
We find that \name allows the sender-chosen policy to be effective:
when configured to implement Stochastic Fairness Queueing (SFQ), it improves median flow completion time (FCT) by between 28\% and 97\% across various scenarios.




\end{abstract}

\maketitle
\setcounter{footnote}{1}
\footnotetext{Both authors contributed equally to this work.}
\renewcommand*{\thefootnote}{\arabic{footnote}}
\setcounter{footnote}{0}
\begin{sloppypar}
\section{Introduction}\label{s:intro}

This paper introduces the idea of {\em site-to-site} Internet traffic control. By ``site'', we mean a single physical location with tens to many thousands of endpoints sharing access links to the rest of the Internet. Examples of sites include a company office, a coworking office building, a university campus, a single datacenter, and a point-of-presence (PoP) of a regional Internet Service Provider (ISP). 


Consider a company site with employees running thousands of concurrent applications. The administrator may wish to enforce certain traffic control policies for the company; for example, ensuring rates and priorities for Zoom sessions, de-prioritizing bulk backup traffic, prioritizing interactive web sessions, and so on. There are two issues that stand in the way: first, the bottleneck for these traffic flows may not be in the company's network, and second, the applications could all be transiting different bottlenecks. So what is the company to do?

Cloud computing has made the second issue manageable. Because the cloud has become the prevalent method to deploy applications today, applications from different vendors often run from a small number of cloud sites (e.g., Amazon, Azure, etc.). This means that the network path used by these multiple applications serving the company's users are likely to share a common bottleneck; for example, all the applications running from Amazon's US-West datacenter, all the video sessions from a given Zoom datacenter, and so on. In this setting,  by treating the traffic between the datacenter site and the company site as a single aggregate, the company's network administrator may be able to achieve their traffic control objectives.


But what about the first issue? The bottleneck for all the traffic between Amazon US-West and the company may not be the site's access link or at Amazon, but elsewhere, e.g., within the company's ISP; indeed, that may be the common case~\cite{inferring-interdomain-congestion, isp-throttle-1, isp-throttle-2, isp-throttle-3}. Unfortunately, the company cannot control traffic when the queues build inside its ISP. And the ISP can't help because it does not know what the company's objectives are.\footnote{Interdomain QoS mechanisms~\cite{braden1997resource, wroclawski1997use} have not succeeded in the Internet despite years of effort.}


\begin{figure*}[ht!]
    \centering
    \includegraphics[width=\textwidth]{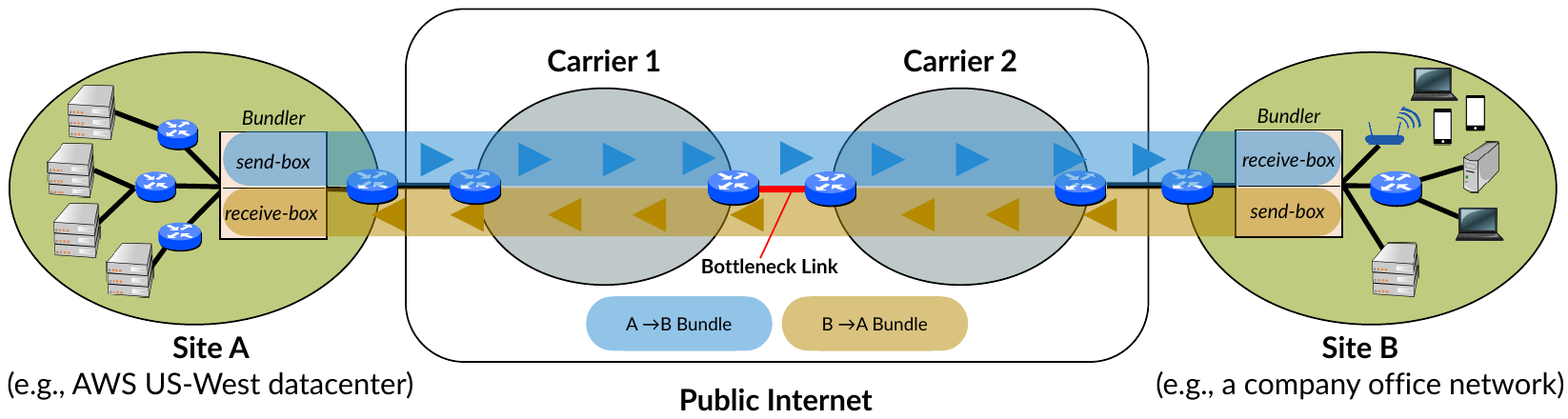}
    \caption{An example deployment scenario for \name in sites A and B.
    Traffic between the two boxes is aggregated into a single bundle, shown as shaded boxes. The \inbox schedules the traffic within the bundle according to the policy the administrator specifies (\S\ref{s:design}).
    }
    \label{fig:deploy:arch}
\end{figure*}

We propose a system, {\em \name}, that solves this problem. \name enables flexible control of a traffic {\em bundle} between a source site and a destination site by {\em shifting} the queues that would otherwise have accumulated elsewhere to the source's site (Figure~\ref{fig:design:shift-bottleneck}) \cut{\radhika{should we instead refer to Fig 2 here?}}. It then schedules packets from this shifted queue using standard techniques~\cite{diffserv, fair-queueing, sfq, pie, CoDel, fifoplus, virtualClocks, csfq, drr, red, ecn} to reduce mean flow-completion times, ensure low packet delays, isolate classes of traffic from each other, etc.



The key idea in \name is a control loop between the source and destination sites to calculate the dynamic rate for the bundle. Rather than terminate end-to-end connections at the sites, we leave them intact and develop an ``inner loop'' control method between the two sites that computes this rate. The inner control loop uses a delay-based congestion control algorithm that ensures high throughput, but controls {\em self-inflicted queueing delays} at the actual bottleneck. By avoiding queues at the bottleneck, the source site can prioritize latency-sensitive applications and allocate rates according to its objectives.

By not terminating the end-to-end connections at the sites, \name{} achieves a key benefit: if the bottleneck congestion is due to other traffic not from the bundle, end-to-end algorithms naturally find their fair-share. It also simplifies the implementation because \name{} does not have to proxy TCP, QUIC, and other end-to-end protocols.

As shown in Figure~\ref{fig:deploy:arch}, \name implements its source site and destination site functions in a \emph{\inbox} and \emph{\outbox}, respectively. The \inbox of one site pairs with the \outbox of another site when sending traffic to it.\footnote{One \inbox can pair with multiple {\outbox}es and vice versa.} 
These two middleboxes measure congestion signals such as the round-trip time (RTT) and the rate at which packets are received, and pass these signals to a congestion control algorithm at the sendbox (\S\ref{s:design}) to dynamically compute the bundle's sending rate.
We introduce a lightweight method for the coordination between the \inbox and the \outbox that does not require any per-flow state and can be deployed in a mode that forwards packets without modification.  \name requires no changes to the end hosts or to network routers.
 
Our focus thus far has been to control traffic only within a given bundle and not across different bundles. 
Furthermore, as we will discuss in \S\ref{s:deploy}, there may be instances where \name cannot improve performance for the bundled traffic, and falls back to the status quo; \ie the performance achieved today when queues build in the network instead of the edge. For example, when traffic between the two sites traverses different paths with different levels of congestion, \name will detect this and performance will revert to the status quo.
 
In emulated scenarios (\S\ref{s:eval}), we demonstrate that \name successfully enables scheduling benefits. In particular, when configured to use Stochastic Fairness Queueing (SFQ),
\name reduces the median flow completion time (FCT) of a representative flow size distribution between 28\% to 97\% across a variety of scenarios. Furthermore, these performance benefits are within 15\% of what would be achievable if (optimal) in-network scheduling were a possibility.
In experiments over the public Internet (\S\ref{s:eval:realworld}), we find that \name reduces short-flow latencies by 57\%.

\section{Related Work}
\label{s:related}


\Para{Traditional congestion control}
End hosts employ congestion control algorithms that aim to achieve high throughput and low delay while fairly sharing network resources with other users~\cite{Jacobson88}. 
Each connection runs such an algorithm independently to learn about the network conditions and find the best sending rate.
In \name, a \inbox uses such an algorithm to determine the aggregate's sending rate, rather than the rate of an individual connection. 
The end-hosts continue to use unmodified end-to-end congestion controllers for each connection.

\Para{Aggregating congestion information} 
There have been multiple proposals to aggregate congestion control information in different contexts: flows sharing the same endpoint~\cite{cm}, flows between two racks within a datacenter~\cite{rackcc}, and flows originating from a large cloud/content provider~\cite{fivecomps}. 
The goal of these approaches is to share information among the various end-to-end flows' congestion controllers, which allows them to better adapt to network conditions. 
\name has a different goal: to control queueing (and thus enable scheduling) from the edge of the network without interfering with the end-to-end congestion controllers of individual component flows. It is orthogonal to prior proposals on aggregate congestion control.

\Para{Using a middlebox for queue management} Remote Active Queue Management (AQM)~\cite{ardelean} aims reduce VoIP traffic latency by deploying a middlebox at a site's access link that drops packets or injects ECN marks for the remaining flows in order to manipulate their end-to-end congestion control loops. It makes a core assumption that the bottleneck is the site's access link.
In contrast, \name tackles arbitrary bottleneck locations in the middle of the network. Moreover, unlike Remote AQM, \name is not restricted to a specific queue management policy for a specific traffic class.



\Para{Overlay networks} \name's motivation is closer to a proposal in overlay networks, OverQoS~\cite{overqos}, which aimed to provide QoS benefits in the Internet by enforcing traffic management policies at the nodes of an already-deployed overlay network~\cite{ron}. 
\name's approach is more lightweight; instead of relying on an overlay network, \name only requires each site to deploy a middlebox, and uses a novel control loop between the middleboxes to facilitate traffic management at the sites. 



\section{Goals and Assumptions}\label{s:deploy}
\begin{figure*}[ht!]
    \centering
    \includegraphics[width=\textwidth]{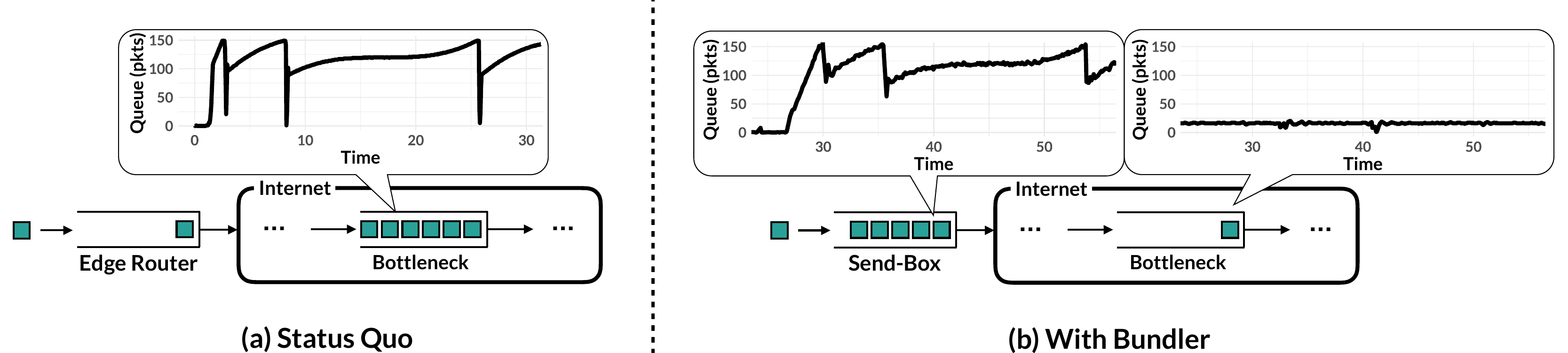}
    \caption{This illustrative example with a single flow shows how \name can take control of queues in the network. The plots, from measurements on an emulated path (as in \S\ref{s:eval}), show the trend in queueing delays at each queue over time. The queue where delays build up is best for scheduling decisions, since it has the most choice between packets to send next. Therefore, the \inbox \emph{shifts} the queues to itself.}\label{fig:design:shift-bottleneck}
\end{figure*}



Figure~\ref{fig:deploy:arch} describes \name's deployment model. 
\name aggregates traffic from Site A to Site B, and vice-versa, into two unidirectional bundles. 
In the egress path, the \inbox moves the in-network queues built by the bundled traffic to itself (illustrated in Figure~\ref{fig:design:shift-bottleneck}) (we describe the specific mechanism in \S\ref{s:design}). 
It can thus enforce desired scheduling policies across the traffic in the bundle.

Our primary goal with \name is to provide control over \emph{self-inflicted} queueing, \ie when traffic from a single bundle causes a queue to build up at the bottleneck links in the network, even without any other cross-traffic.
In the remainder of this section, we detail the conditions in which \name can achieve this goal. Our high-level strategy is ``do no harm''; \name detects conditions in which it cannot operate and temporarily disables itself until favorable conditions return, reverting to status-quo performance in the meantime.

\Para{Non-edge bottleneck}
Network administrators already have control over packets which queue within their site. \name is capable of taking control over queues that build up anywhere between a \pair. Thus, deploying \name at the edge of each site captures any potential build up outside of either site's control. Such congestion might occur at an inter-domain link, within either site's ISP, or, if a site is managed by a cloud provider, it could even occur within its datacenter (\eg{} at the cloud provider's rate limiter, \S\ref{s:eval:realworld}). 

There is strong evidence that such non-edge bottlenecks exist.
Dhamdhere \etal measured~\cite{inferring-interdomain-congestion} inter-domain bottlenecks such as the red bottleneck link in Figure~\ref{fig:deploy:arch}.
Similarly, Zhu \etal found~\cite{bottleneck-of-china} that non-edge bottlenecks for transnational traffic to and from China are prevalent, and moreover that in many cases, the bottleneck for this traffic is an ISP deep inside China rather than a larger provider.
An M-Lab technical report similarly found~\cite{mlab-tr} patterns of performance degradation linked to specific ISP interconnections in the middle of the network.
Finally, Jin \etal found~\cite{blameit} that for WAN traffic originating from Microsoft Azure, the ``middle'', \ie on-path ASes not including the source or destination AS, is to blame for between $40$-$50\%$ of persistent congestion incidents over a one-month period. 

\Para{External congestion} Other than self-inflicted congestion, \name must coexist with traffic from external sources.

\vspace{2pt}
\paragraphi{Congestion due to bundled cross-traffic}
\name continues to provide benefits when the competing flows are part of other bundles from/to other sites because the rate control algorithm at each of the other {\inbox}es would ensure that the in-network queues remain small, and different bundles compete fairly with one another. Since each \inbox manages the self-inflicted queues for its own bundles, it can apply the appropriate scheduling policy in its per-bundle queues.

\vspace{2pt}
\paragraphi{Congestion due to un-bundled cross-traffic} We now consider the scenario where the cross-traffic includes un-bundled flows. If all such \emph{un-bundled} competing flows are short-lived (up to a few MBs) or application-limited (\eg a paced video stream), the bundled traffic still sees significant performance benefits, because there are not enough packets in such short-lived flows to fill up the queues or claim a greater share of network bandwidth.
However, if the cross traffic is long lasting, and employs a loss-based congestion controller to send back-logged bulk data, it aggressively fills up all available buffer space at the bottleneck link. Naively using a delay-based congestion controller at \name against such aggressive \emph{buffer-filling} cross-traffic would severely degrade the throughput of the bundled traffic. Therefore, \name's congestion controller detects the presence of buffer-filling cross-traffic;
to compete fairly, it relinquishes most of its control (and scheduling opportunities) over the bundled traffic, while still maintaining a small queue for continued detection of cross-traffic (detailed in \S\ref{s:buffer-filling}). 
However, such pathological buffer-filling cross traffic is rare.
A recent study in CDNs~\cite{akamai-cdn-trace} and our analysis of a packet trace from an Internet backbone router~\cite{caida-dataset} reveal that the vast majority of connections are smaller than 1MB: too small to build persistent queues.\footnote{\cut{\radhika{check:} }This implies that flows \emph{within} a bundle may also be short-lived requests or paced audio/video traffic which, when aggregated by \name, can form a heavy, long-lasting bundle.} 
Our experiments on Internet paths (\S\ref{s:eval:realworld}), also did not encounter pathological buffer-filling cross traffic.  


\Para{Shared congestion across flows in the bundle} Bundler’s design for moving queues via aggregate congestion control assumes that the component flows within a bundle share in-network paths, and thus congestion. 
To test this assumption, we used Scamper~\cite{scamper} to probe all paths to $5000$ random IPv4 addresses from each of $30$ cloud instances across the regions of public cloud providers AWS and Azure. In no cases did we find that probe packets took different AS-level paths through the network.
However, we observed instances of IP-level load balancing in 26\% of IP hops. 
In pathological scenarios with \emph{persistent imbalance} in queueing between the load-balanced paths, \name cannot gather accurate measurements and perform aggregated delay-based rate control for the bundled traffic. 
Designing a new congestion control algorithm for such scenarios remains an avenue for future work.
Nonetheless, \name can detect these scenarios (\S\ref{s:queue-ctl:ecmp}) and disable its rate control in such cases, falling back to status-quo performance. 
We expect a well-implemented load balancer will work to prevent persistent imbalance from occurring; indeed, our success with using \name on real Internet paths (\S\ref{s:eval:realworld}) suggests that such pathological cases do not occur in practice.

\paragrapha{Intuition for \name's applicability}
Another way of understanding when \name is useful, which incorporates the three conditions above, is the following litmus test:
compare the queueing delay when all flows (\name's and cross-traffic) are in the network with the queueing delay when the \name’s flows are magically removed; if the latter is lower than the former, then \name can provide benefits.

\cut{
\vspace{0.05in}
\paragrapha{Takeaway} While \name must yield its scheduling ability in the face of aggressive, buffer-filling cross traffic and persistently imbalanced load balancing, in most scenarios it significantly improves performance (as we show in \S\ref{s:eval}).
This, combined with its deployment ease, makes a strong case for deploying \name. 
}

\section{Designing \name}\label{s:design}
\cut{
\an{maybe move to ``bundler's utility regime''/Overview/``Traffic Bundles'': 
What is the best way to take advantage of the existence of traffic bundles?
Broadly, there are two approaches: congestion control and in-network scheduling. 
Today, congestion control is a distributed concern: each end-host performs its own optimization to achieve the highest throughput and lowest latency.
Indeed, end-host rate control is a deployed method to enforce domain-wide scheduling policy in private WANs~\cite{bwe}.
On the other hand, scheduling is an in-network concern; existing proposals such as DiffServ~\cite{diffserv} observe that the network is uniquely positioned to differentiate between different flows according to some domain-wide policy.
Scheduling presents its own challenge: because the Internet is a federated system, the link with queue build-up (and thus with the opportunity to re-order packets according to a scheduling policy) will likely be outside the sending domain's purview.

\an{don't know what to say next}
}
}

Recall that in order to do scheduling, we need to move the queues from the network to the \name. 
In this section, we first describe our key insight for moving the in-network queues, and then explain our specific design choices. 
Recall that each site deploys one \name middlebox which we logically partition into sender-side (\inbox) and receiver-side (\outbox) functionality.

\subsection{Key Insight}\label{s:design:key}
We induce queuing at the \inbox by rate limiting the outgoing traffic. 
If this rate limit is made smaller than the bundle's fair share of bandwidth at the bottleneck link in the network, it will decrease throughput. 
Conversely, if the rate is too high, packets will pass through the \inbox without queueing.
Instead, the rate needs to be set such that the bottleneck link sees a small queue while remaining fully utilized (and the bundled traffic competes fairly in the presence of cross traffic). 
We make a simple, but powerful, observation: existing congestion control algorithms calculate exactly this rate~\cite{Jacobson88}. 
Therefore, running such an algorithm to set a bundle's rate would reduce its self-inflicted queue at the bottleneck, causing packets to queue at the \inbox instead, without reducing the bundle's throughput.
Note that end hosts would continue running a traditional congestion control algorithm as before (\eg Cubic~\cite{cubic}, BBR~\cite{bbr}) which is unaware of \name.
Rather, the \inbox's congestion control algorithm acts on the traffic bundle as a \emph{single unit}.

Figure~\ref{fig:design:shift-bottleneck} illustrates this concept for a single flow traversing a bottleneck link in the 
network.\footnote{Details of the emulated network setup which resulted in the illustrated queue length time-series are in \S\ref{s:eval}.}
Without \name, packets from the end hosts are queued in the network, while the queue at the edge is unoccupied. 
In contrast, a \name deployed at the edge is able to shift the queue to its \inbox.

\subsection{System Overview}\label{s:design:overview}
\begin{figure}
    \centering
    \includegraphics[width=\columnwidth]{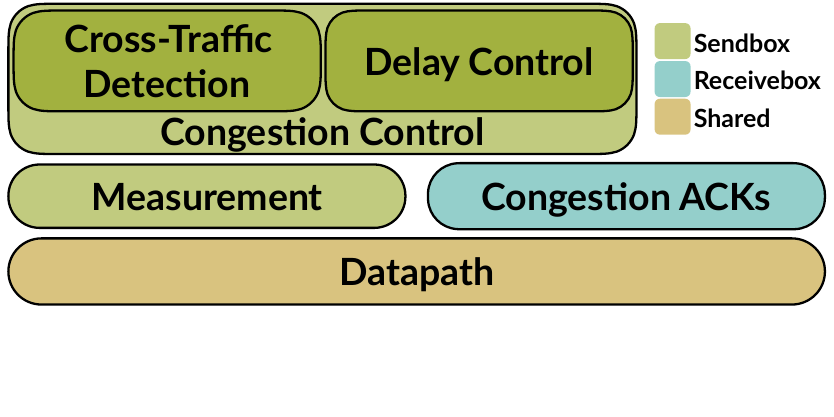}
    \vspace{-40pt}
    \caption{\name comprises of six sub-systems: four (in green) implement \inbox functionality, one (in blue) implements \outbox functionality, and the datapath (orange) is shared between the two. \cut{\radhika{should we put the \name blocks inside another box called ``control plane''?}}}\label{fig:design:block-diag}
\end{figure}
Figure~\ref{fig:design:block-diag} shows \name's sub-systems: 
(1) A congestion control module at the \inbox which implements the rate control logic and cross-traffic detection, as discussed in \S\ref{s:design:whichcc}.
(2) A mechanism for sending congestion feedback (ACKs) in the \outbox, and (3) a measurement module in the \inbox that computes congestion signals (RTT and receive rate) from the received feedback. We discuss options for implementing congestion feedback mechanism in \S\ref{s:design:twosided} and how to use that feedback in the measurement module in \S\ref{s:measurement}.
(4) A datapath for packet processing (which includes rate enforcement and packet scheduling). Any modern middlebox datapath, \eg BESS~\cite{bess}, P4~\cite{p4}, or  Linux qdiscs (as used in our prototype implementation---see \S\ref{s:impl}), is suitable. 
We detail the interaction between these subsystems when discussing our prototype implementation in \S\ref{s:impl}. 
%
%
In the rest of this section, we discuss our key design choices.

\subsection{Choice of congestion control algorithm}\label{s:design:whichcc}
\name's congestion control algorithm must satisfy the following requirements: 

\paragraphi{(1) Ability to limit network queueing} \name must limit queueing in the network to move the queues to the \inbox. Therefore, congestion control algorithms which are designed to control delays, and thus queueing, are the appropriate choice. 
A loss-based congestion control algorithm which fills buffers (\eg Cubic, NewReno), for example, is not a good choice for \name, since it would build up a queue at the network bottleneck and drain queues at the \inbox.

\paragraphi{(2) Detection of buffer-filling cross-traffic} It is well known that delay-controlling schemes (\eg Vegas~\cite{vegas}) compete poorly with buffer-filling loss-based schemes~\cite{copa}.
Therefore, \name must have a mechanism to detect the presence of such competing buffer-filling flows and fall back to status quo performance, and then detect when they have left to take back its control over the network queues. 

The emergence of such detection mechanisms is recent: Copa~\cite{copa} detects whether it is able to empty the queues, and Nimbus~\cite{nimbus-arxiv} provides a more general mechanism which overlays a pattern on the sending rate and measures the cross traffic's response.
Copa is not designed for aggregate congestion control (see \S\ref{s:queue-ctl}); thus, we use the more general Nimbus mechanism.

\subsection{Congestion Feedback Mechanism}\label{s:design:twosided}
A congestion control algorithm at the \inbox would require network feedback from the receivers to measure congestion and adjust the sending rates accordingly. We discuss multiple options for obtaining this. 


\paragrapha{Passively observe in-band TCP acknowledgements}
Conventional endhost-based implementations have used TCP acknowledgements to gather congestion control measurements. A simple strategy for \name is to passively observe the receiver generated TCP acknowledgements at the \inbox. However, we discard this option as it is specific to TCP and thus incompatible with alternate protocols, \ie UDP for video streaming or QUIC's encrypted transport header~\cite{quic}.

\paragrapha{Terminate connections and proxy through TCP} With this approach, one would terminate end-host TCP connections at the \inbox and open new connections to the \outbox, allowing the \inbox to control the rate of traffic in these connections.
This approach can improve performance by allowing end-to-end connections to ramp up their sending rates quickly. 
The primary disadvantage of this approach is that \name must take responsibility for reliable delivery of component traffic, which requires large amounts of queueing and, in the case of UDP applications, can harm application performance. 
Furthermore, proxying TCP connections introduces a new potential point of failure at \name that violates fate-sharing; if \name crashes, connections will be lost.
Finally, from a practical standpoint, to avoid head-of-line blocking this approach requires that \name open a new proxy connection for each component end-host connection, but still determine the bottleneck rate of the traffic \emph{aggregate}. While this approach may be technically feasible~\cite{cm}, it would result in high overhead.
\cut{\footnote{From an architectural standpoint this design runs counter to the end-to-end principle~\cite{e2e-principle}; it replicates endhost functionality in the network.}}
Thus, we set aside TCP proxies for the remainder of this discussion, but explore their compatibility with \name in \S\ref{s:eval:proxy}. 

\paragrapha{Out-of-band feedback} Having eliminated the options for using in-band feedback, we adopt an out-of-band feedback mechanism: the \outbox sends out-of-band congestion ACKs to the \inbox.
This decouples congestion signalling from traditional ACKs used for reliability and is thus indifferent to the underlying protocol (be it TCP, UDP, or QUIC).

\cut{
\begin{Appendix}
\section{Nimbus}\label{s:app:nimbus}
\begin{figure*}[ht!]
    \centering
\begin{knitrout}
\definecolor{shadecolor}{rgb}{0.969, 0.969, 0.969}\color{fgcolor}
\includegraphics[width=\maxwidth]{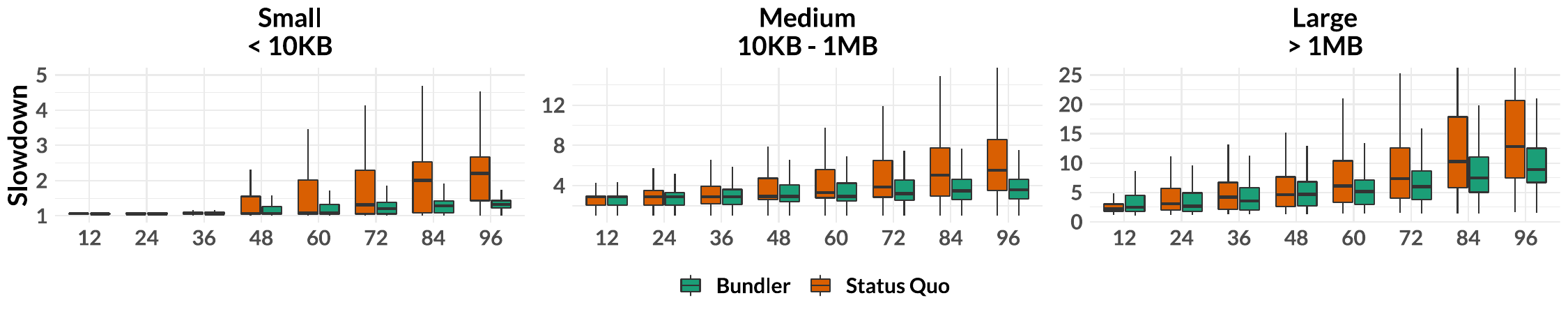} 

\end{knitrout}
    \caption{\name offers diminishing returns with lower amounts of offered load.}
    \label{fig:eval:offeredload}
\end{figure*}

We briefly summarize Nimbus~\cite{nimbus} for reference.

Nimbus's goal is to detect scenarios in which it is safe to use delay-based congestion control. 
To achieve this goal, Nimbus proposes an \emph{elasticity detector} to detect whether the cross traffic contains any \emph{elastic} flows, which react to changes in the available bandwidth on fast time-scales, \ie a couple RTTs. 
Nimbus's core observation is that the absence of such elastic cross traffic, \ie competition with only \emph{inelastic} traffic, is a sufficient (but not necessary) condition to use a delay-control algorithm.
Nimbus thus measures whether cross traffic reacts to changes in available bandwidth by pulsing its sending rate at a predetermined frequency.
If the cross traffic's rate responds to these pulses at the same frequency, Nimbus can conclude that it is elastic, because it has reacted to changes in the available bandwidth.

A natural method of measuring the cross traffic's frequency response is to compare Nimbus's pulses with the cross traffic's rate in the frequency domain.
Thus, Nimbus uses an asymmetric sinusoidal pulse (as we have noted in \S\ref{s:queue-ctl}) which has a straightforward representation in the frequency domain while maximizing the sender's ability to influence the available bandwidth.

How can Nimbus know the cross traffic's response? It develops an estimator for the cross traffic's sending rate:
\begin{equation}
    \hat{z}(t) = \mu\frac{S(t)}{R(t)} - S(t)
\end{equation}

$\hat{z}(t)i$ is the estimated cross traffic rate, $\mu$ is the estimated bottleneck bandwidth, $S(t)$ is the sending rate, and $R(t)$ is the receiving rate. 
    \name measures $R(t)$ using congestion ACKs from the \outbox and $\mu$ using the maximum receive rate as in BBR~\cite{bbr}.

Then, Nimbus searches for a peak in the neighborhood of its pulsing frequency $f_p$ to determine the elasticity metric $\eta$:

\begin{equation}
    \eta = \frac{|\text{FFT}_{\hat{z}}(f_p)|}{\text{max}_{f \in (f_p, 2f_p)} |\text{FFT}_{\hat{z}}(f)|}
\end{equation}

If $\eta$ is larger, the cross traffic is more elastic.
Nimbus then uses a hard-decision rule $\eta \ge 2$ to decide when to switch to cross-traffic competitive mode as described in \S\ref{s:queue-ctl}.

\end{Appendix}
}

\subsection{Measuring Congestion}\label{s:measurement}
\begin{figure}
    \centering
    \includegraphics[width=\columnwidth]{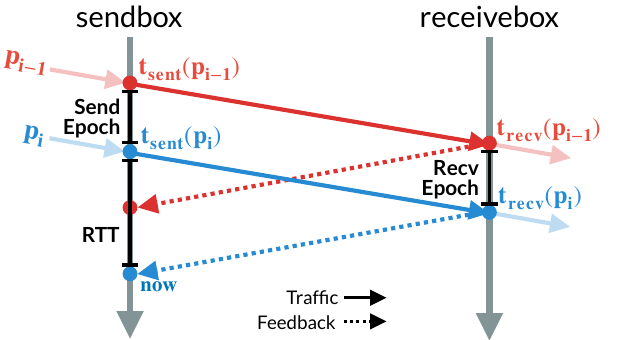}
    \caption{Example of epoch-based measurement calculation. Time moves from top to bottom.
    The \inbox records the packets that are identified as epoch boundaries. 
    The \outbox, up on identifying such packets, sends a feedback message back to
    the \inbox, which allows it to calculate the RTT and epochs.
    }\label{fig:ratecalc}
\end{figure}



Sending an out-of-band feedback message for every packet arriving at the \outbox would result in high communication overhead. 
Furthermore, conducting measurements on every outgoing packet at the \inbox would require maintaining state for each of them, which can be expensive, especially at high bandwidth-delay products. 
This overhead is unnecessary; reacting once per RTT is sufficient for congestion control algorithms~\cite{ccp}. 
The \inbox therefore samples a subset of the packets for which the \outbox sends congestion ACKs.
We refer to the period between two successively sampled packets as an \emph{epoch}, and each sampled packet as an \emph{epoch boundary packet}.

The simplest way to sample an epoch boundary packet would be for the \inbox to probabilistically modify a packet (\ie set a flag bit in the packet header) and the \outbox to match on this flag bit.
However, where in the header should this flag bit be?
Evolving packet headers has proved impractical~\cite{trotsky}, so perhaps we could use an encapsulation mechanism.
Protocols at both L3 (\eg NVGRE~\cite{nvgre}, IP-in-IP~\cite{ipinip}) and L4 (\eg VXLAN~\cite{vxlan}) are broadly available and deployed in commodity routers today.

Happily, we observe that such packet modification is not inherently necessary; since the same packets pass through the \inbox and \outbox, uniquely identifying a given pattern of packets is sufficient to meet our requirements. In this scheme, the \inbox and \outbox both hash a subset of the header for every packet, and consider a packet as an epoch boundary if its hash is a multiple of the desired \emph{sampling period}.

Upon identifying a packet $p_i$ as an epoch boundary packet the \inbox records: 
(i) its hash, $h(p_i)$, 
(ii) the time when it is sent out, $t_{\text{sent}}(p_i)$, 
and (iii) the total number of bytes sent thus far including this packet, $b_{sent}(p_i)$. 
When the \outbox sees $p_i$, it also identifies it as an epoch boundary and sends a congestion ACK back to the \inbox. 
The congestion ACK contains $h(p_i)$ and the running count of the total number of bytes received for that bundle. 
Upon receiving the congestion ACK for $p_i$, the \inbox records the received information, and using its previously recorded state, computes the RTT and the rates at which packets are sent and received, as in Figure~\ref{fig:ratecalc}.

\begin{figure}[t]
    \centering
\begin{knitrout}
\definecolor{shadecolor}{rgb}{0.969, 0.969, 0.969}\color{fgcolor}
\includegraphics[width=\maxwidth]{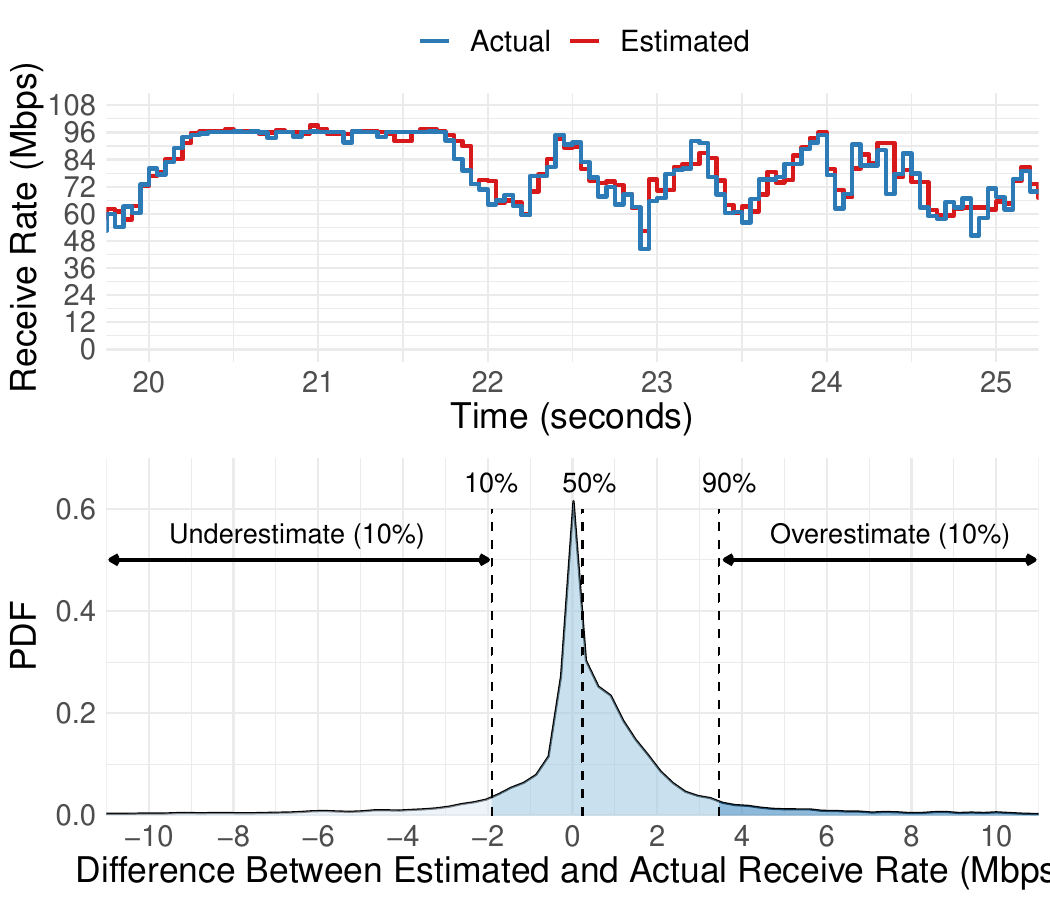} 

\end{knitrout}

    \caption{\name's estimate of the receive rate.}
    \label{fig:micro:time-thru}
\end{figure}

\paragrapha{Epoch boundary identification}
The packet header subset that is used for identifying epoch boundaries must have the following properties:
(i) It must be the same at both the \inbox and the \outbox.
(ii) Its values must remain unchanged as a packet traverses the network from the \inbox to the \outbox (so, for example, the TTL field must be excluded).\footnote{Certain fields, that are otherwise unchanged within the network, can be changed by NATs deployed within a site. Ensuring that the \name boxes sit outside the NAT would allow them to make use of those fields.}
(iii) It differentiates individual \emph{packets} (and not just flows), to allow sufficient entropy in the computed hash values.
(iv) It also differentiates a retransmitted packet from the original one, to prevent spurious samples from disrupting the measurements (this precludes, for example, the use of TCP sequence number).
We expect that the precise set of fields used will depend on specific deployment considerations.
For example, in our prototype implementation (\S\ref{s:impl}) we use a header subset of the IPv4 IP ID field and destination IP and port. 
We make this choice for simplicity; it does not require tunnelling mechanisms and is thus easily deployable, and if \name fails, connections are unaffected. 
We note that previous proposals~\cite{ip-traceback} have used IP ID for unique packet identification. 
The drawback of this approach is that it cannot be extended to IPv6.
To support a wider set of scenarios, \name could use dedicated fields in an encapsulating header (as in~\cite{axe}).

To visualize how these measurements impact the behavior of the signals over time we pick an experiment for which the median difference matches that of the entire distribution and plot a five second segment of our estimates compared to the actual values in Figure~\ref{fig:micro:time-thru}.

\begin{figure}[t]
    \centering
\begin{knitrout}
\definecolor{shadecolor}{rgb}{0.969, 0.969, 0.969}\color{fgcolor}
\includegraphics[width=\maxwidth]{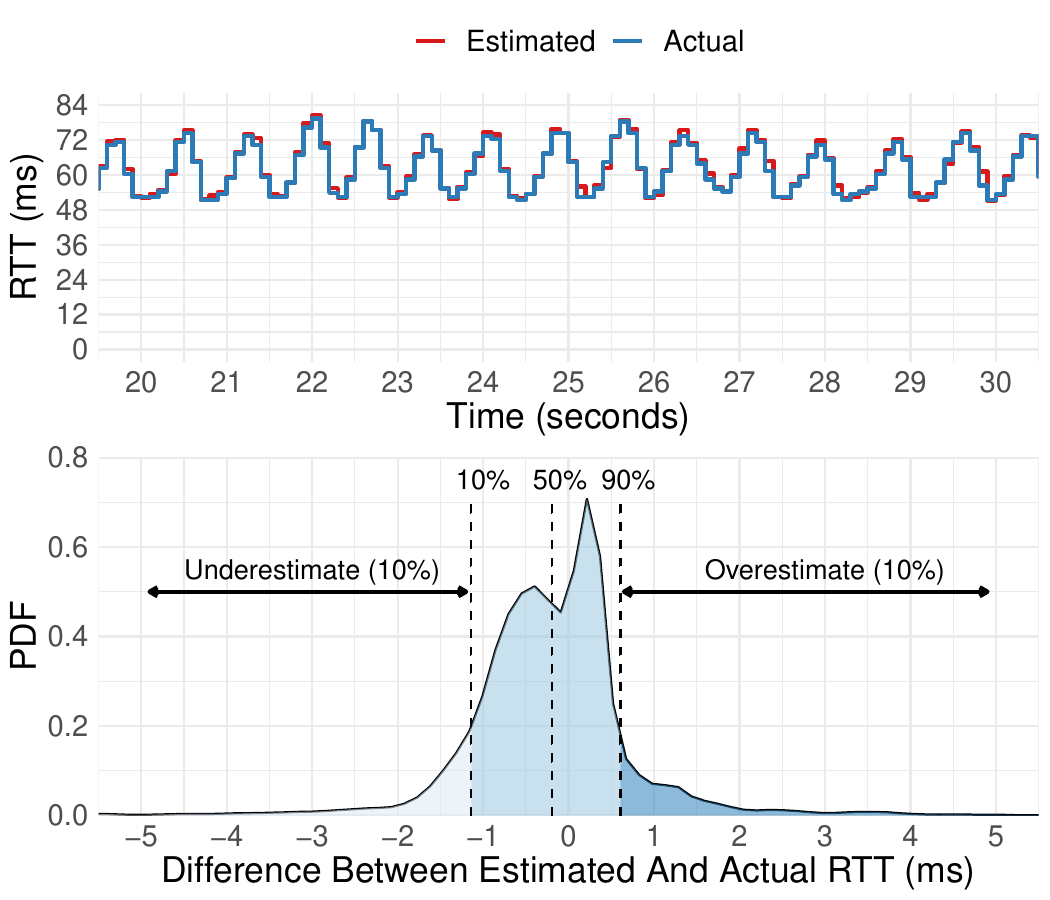} 

\end{knitrout}

    \caption{\name's estimate of the delay }
    \label{fig:micro:time-delay}
\end{figure}

\paragrapha{Choosing the epoch size}
In order to balance reaction speed and overhead, epoch packets should be spaced such that measurements are collected approximately once per RTT~\cite{ccp}.
Therefore, for each bundle, we track the minimum observed RTT ($minRTT$) at the \inbox and set the epoch size $N = (0.25 \times minRTT \times send\_rate)$, where the $send\_rate$ is computed as described above. The measurements passed to the congestion control algorithms at the \inbox are then computed over a sliding window of epochs that corresponds to one RTT. Averaging over a window of multiple epochs also increases resilience to possible re-ordering of packets between the \inbox and the \outbox, which can result in them seeing different number of packets between two epochs.

When the \inbox updates the epoch size $N$ for a bundle, it needs to send an out-of-band message to the \outbox communicating the new value. To keep our measurement technique resilient to potential delay and loss of this message, the epoch size $N$ is always rounded down to the nearest power of two. Doing this ensures that the epoch boundary packets sampled by the \outbox are either a strict superset or a strict subset of those sampled by the \inbox. The \inbox simply ignores the additional feedback messages in former case, and the recorded epoch boundaries for which no feedback has arrived in the latter.  

\paragrapha{Robust to packet loss}
Note that our congestion measurement technique is robust to a boundary packet being lost between the \inbox and the \outbox. In this case, the \inbox would not get feedback for the lost boundary packet, and it would simply compute rates for the next boundary packet over a longer epoch once the next congestion ACK arrives.

\paragrapha{Microbenchmarks}
To evaluate the accuracy and robustness of this measurement technique, we picked 90 traces from our evaluation covering a range of link delays (20ms, 50ms, 100ms) and bottleneck rates (24Mbps, 48Mbps, 96Mbps), and computed the difference, at each time step, between \name's measurements (estimate) and the corresponding values measured at the bottleneck router (actual). 
In Figure~\ref{fig:micro:time-delay} we focus on the RTT measurements: the bottom plot shows the distribution of the differences, and the top plot puts it into context by showing a five second segment from a trace where the median difference matched that of the full distribution. In Figure~\ref{fig:micro:time-thru}, we produce the same plots for the receive rate estimates.
In summary, 80\% of our RTT estimates were within 1.2ms of the actual value, and 80\% of our receive rate estimates were within 4Mbps of the actual value.

\subsection{Implications of \name's Design}

Our design choices result in an architecture where \name's inner rate control loop can be implemented entirely the ``control plane'' of the \inbox and the \outbox, which passively observes the packets traversing the datapath of the middleboxes, without modifying them. This results in a truly transparent system, that is light-weight, has low overhead, preserves fate-sharing, and in no way interferes with the end-to-end controllers of individual flows. The only datapath action that \name performs is the enforcement of the desired scheduling and queue management policies at the \inbox. 

\section{\cut{Handling }Unfavorable Conditions}\label{s:queue-ctl}

Recall from \S\ref{s:deploy} that \name can reliably shift queue build up from the bottleneck to itself when, (a) the cross-traffic is not buffer-filling, and (b) all of its component traffic shares the same bottleneck in the network.
In practice, either of these conditions may break. 
In this section, we describe how \name can re-use the same measurements from \S\ref{s:measurement} to detect when these conditions do not hold. In such cases, \name (temporarily) disables its rate limiting (falling back to status-quo performance) until favorable conditions arise again. 

\subsection{Buffer-Filling Cross Traffic}
\label{s:buffer-filling}


It is well known that delay-based congestion control algorithms (as \name uses) lose throughput when competing with buffer-filling algorithms~\cite{copa, nimbus-arxiv}. 
To prevent this, \name utilizes prior work, Nimbus~\cite{nimbus-arxiv}, which provides a mechanism for detecting the presence of buffer-filling\footnote{In particular, Nimbus detects ``elastic'' cross-traffic~\cite{nimbus-arxiv}, a superset of buffer-filling traffic. \cite{nimbus-arxiv} provides an explanation of this distinction
and a detailed evaluation of Nimbus' accuracy of detecting elastic cross traffic and speed of switching between the two modes, using both emulated and real-world experiments. \name{}'s use of Nimbus does not impact its accuracy or speed of switching.} 
cross traffic, and proposes temporarily switching to a buffer-filling scheme to compete fairly whenever such cross traffic is present.
At a high-level, the detection mechanism works as follows: given a desired sending rate $r(t)$ (from an underlying congestion control algorithm), Nimbus superimposes an asymmetric sinusoid onto $r(t)$ to determine the sending rate. Then, it monitors the measured send and receive rate, estimates the cross-traffic's rate, and monitors the cross traffic's rate in the frequency domain. The sinusoidal variations in the sending rate will be visible in the cross-traffic's rate only if buffer-filling cross traffic is sharing the same bottleneck queue. 

What exactly should the \inbox do when it detects buffer-filling traffic? Using a buffer-filling scheme for the bundle as in Nimbus would be fraught: since a bundle is comprised of many individual flows, the \inbox would need to know the number of flows in the bundle to know how aggressively it should compete in order to receive its fair share (as in the status quo)~\cite{multcp}. 
This number may vary significantly over time and would be difficult to measure, especially on high-performance datapaths~\cite{heavy-hitters}.

Instead, we propose a simpler solution.
Since each connection in a bundle is already employing its own congestion controller, \name{} can simply \emph{let the traffic pass}, \ie{} increase the pacing rate at the \inbox to stop controlling queues.
Then, the end-host congestion control loops will naturally compete fairly with the buffer-filling cross traffic, just as they would without \name{}.

However, letting the traffic pass creates a new challenge. 
To determine when it is safe to resume delay-control while in the traffic-passing mode, Nimbus requires a superimposed pulse in both modes. If we naively let the traffic pass, the \inbox queue would never build. As a result, there would not be sufficient packets in the queue to perform the rate increase for the up-pulse. 
Without the up-pulse, once the \inbox{} switched to the buffer-filling mode, it would not be able to gather sufficient information to switch back to delay-control mode once the buffer-filling cross traffic subsided.

To support the Nimbus pulses while also letting the traffic pass,
the \inbox must maintain sufficient queueing for the up-pulse,
\ie the area under the up-pulse curve: 
$A \int_0^{\frac{T}{4}} \sin(\frac{4\pi{}t}{T}) dt = \frac{AT}{2\pi}$.
From Nimbus, we use $T = 0.2$ seconds and $A =$ one-fourth the bottleneck bandwidth ($\mu$),
which yields
$\frac{T\mu}{8\pi}$, or $8\text{ms}\cdot \mu$ of queueing. 
We thus configure the \inbox to maintain a target queue $q_T$ of $10$ms (the additional $2$ms is a cushion against input variance).
Because bundled connections will experience this queueing in addition to other queueing in the network, 
most traditional congestion control algorithms (\eg Cubic) will observe RTT inflation. 
In \S\ref{s:robust:cross} we show that this effect is not large; \name still achieves performance comparable to the status quo. Nevertheless, it is desirable minimize this inflation and be as close to $q_T$ as possible. 

Thus, to achieve the target queue $q_T$
we use a PI controller at the \inbox which determines how the base sending rate $r(t)$ should be updated:
$\dot{r}(t) = \alpha (q(t) - q_T) + \beta (\dot{q}(t))$, where $\dot{r}(t)$ is the update to $r(t)$ before imposing the Nimbus pulse, $q(t)$ is the current queue size at the \inbox, and $\alpha$ and $\beta$ are both positive.
If $q(t) > q_T$, the first term will be positive and the rate will increase, causing the queue to shrink. Similarly, if the queue size is growing, the second term will be positive, which means the rate will increase and the queue will shrink. 
Setting $\alpha$ and $\beta$ controls a tradeoff: with larger values the controller will approach the target faster, but if they are too large the controller's variations will dominate the Nimbus pulse. 
If they are too small, it will take too long to reach the target. 
We found that $\alpha=10$ and $\beta = 10$ work well for the scenarios in our evaluation (\S\ref{s:eval} and \S\ref{s:eval:realworld}).


\subsection{Imbalanced Multipathing}\label{s:queue-ctl:ecmp}
\begin{figure}
    \centering
\includegraphics[width=\maxwidth]{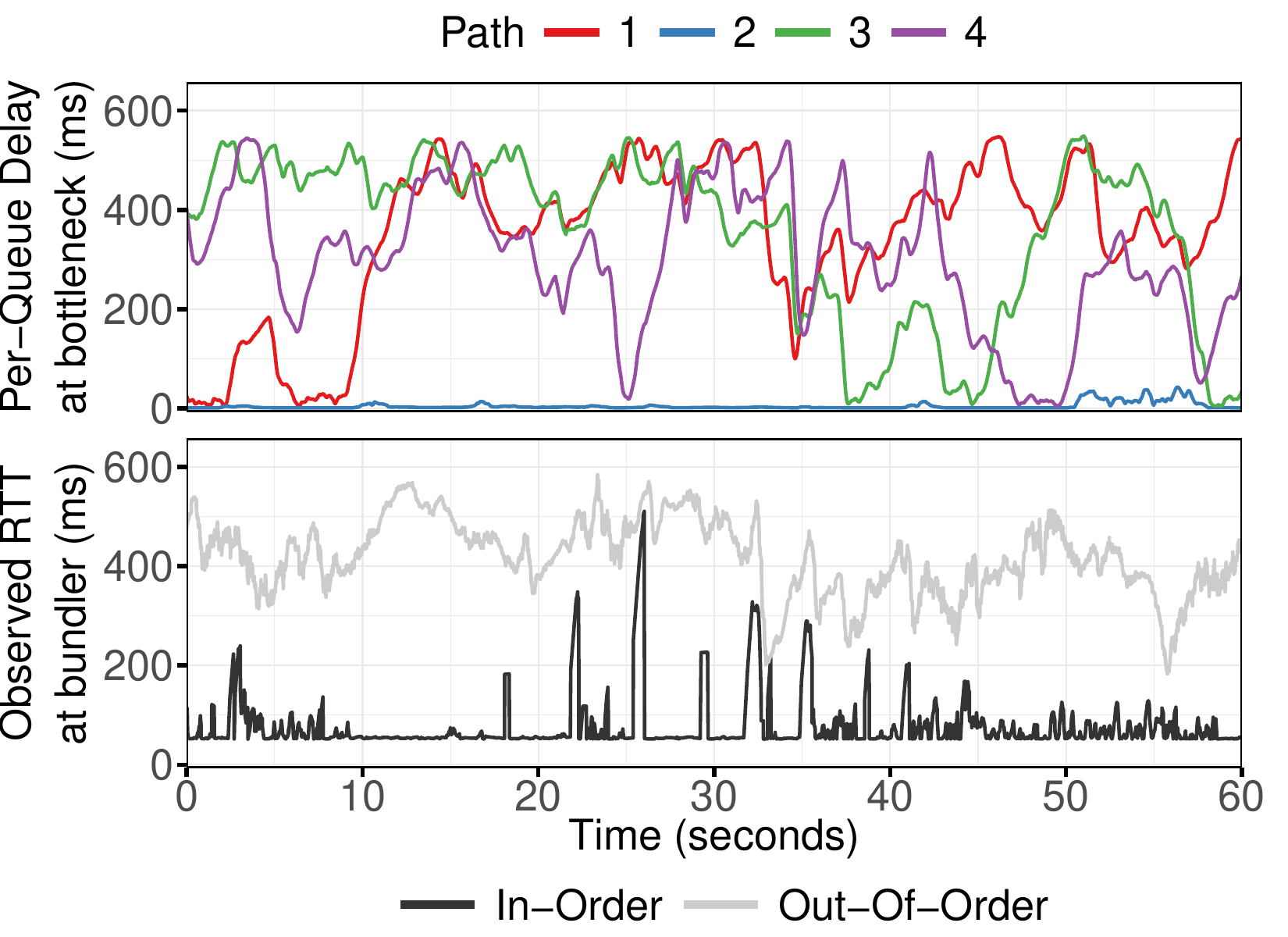} 
\vspace{2pt}
\caption{(Top) True delay for all packets of \name's component flows based on which of 4 load-balanced paths they traversed (unknown to \name). (Bottom) Delay measurements observed by Bundler, colored based on whether they were derived from an in-order or out-of-order epoch packet. Bundler's measurements cannot distinguish how many paths there are, but the relative number of out-of-order measurements is enough to clearly indicate the presence of multiple RTT-imbalanced paths.}
\label{fig:queue-ctl:ecmp:motivation}
\end{figure}

Since a \bundle contains many component connections, a load balancer may send them along different paths. If the load along different paths is well-balanced, \name will accurately treat a load-balanced bottleneck link as a single link whose rate is the sum of the rates of each sub-link. However, when the load along different paths is imbalanced, the series of measurements \name collects will be a random sampling of the different paths, which would confuse the delay-control algorithm and cause it to perform poorly. Fortunately, such cases are straight-forward to detect with our measurement technique. 
More specifically, load imbalance will result in many epoch measure packets arriving out-of-order at the \outbox (whenever epoch packet $i$ happens to traverse a path with a larger delay than epoch packet $i+1$), and consequently, out-of-order ``congestion ACKs'' at the \inbox.  Figure~\ref{fig:queue-ctl:ecmp:motivation} demonstrates this in an emulated imbalance scenario. 

Therefore, we use the fraction of epoch measurement packets that arrive out-of-order as an indicator of load imbalance due to multipathing.
If this number is small, the links are roughly balanced and \name will operate as expected.
If it is large, it indicates load imbalance, in which case \name's rate control may not work well. 
In \S\ref{s:eval:ecmp}, we experimentally determine an out-of-order fraction of 5\% to be a good threshold indicating whether or not the links are balanced: all single-path scenarios resulted in an order of magnitude fewer out-of-order packets, and all multi-path scenarios resulted in an order of magnitude greater.


\section{Implementation}\label{s:impl}
\begin{figure*}[t]
    \centering
    \includegraphics[width=2\columnwidth]{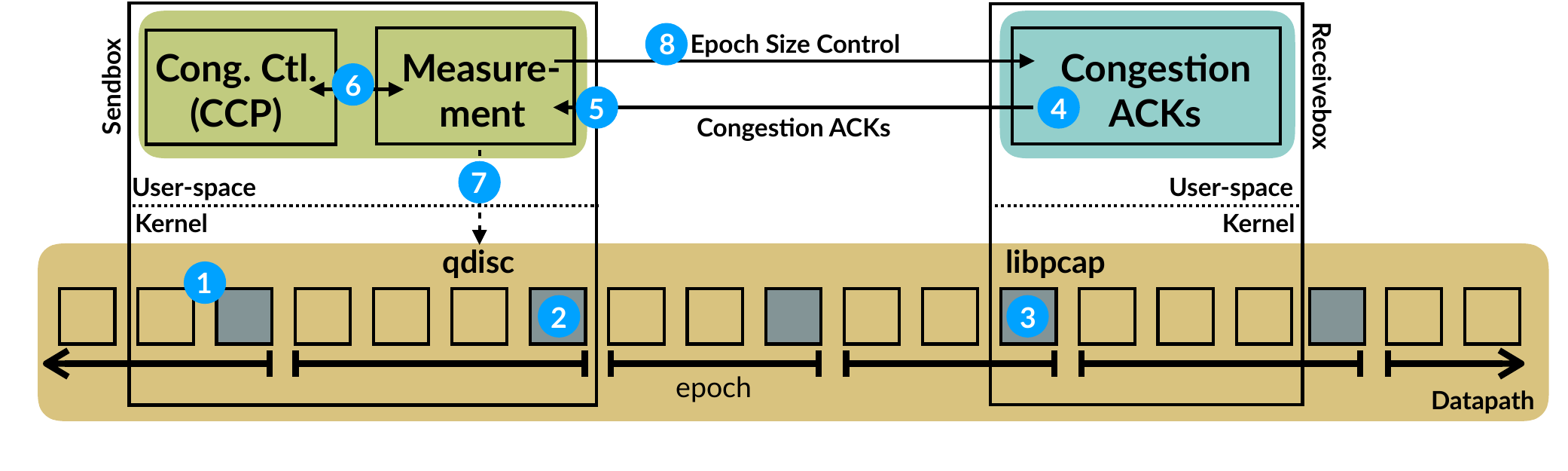}
    \vspace{-10pt}
    \caption{\name Implementation Overview. Box colors correspond to the roles described in Figure~\ref{fig:design:block-diag}. Shaded packets are those that meet the epoch boundary condition. Dashed arrows represent communication via IPC, while solid arrows represent communication over the network.}\label{fig:bundler}
\end{figure*}
\name boxes can be implemented as described below (although the specific implementations could vary across deployments).

\Para{\capinbox} It comprises of a \emph{data plane} and a \emph{control plane}. The data plane is responsible for (i) packet forwarding, (ii) tracking the number of sent bytes, (iii) identifying and reporting the epoch boundary packets to the control plane, (iv) enforcing a sending rate (computed by the control plane) on a bundle, and (iv) enforcing the desired scheduling policies for a bundle. 
It can implemented in software~\cite{bess, click, netbricks, tc}, or in programmable hardware~\cite{p4}. 
The control plane, implemented in software, is responsible for (i) measuring congestion signals using the information provided by the data plane along with the feedback from the \outbox, (ii) computing and communicating epoch sizes, and (iii) running the congestion control algorithm for each bundle to compute appropriate sending rates based on the measured congestion signals.

\Para{\capoutbox} It (i) tracks the number of received bytes, (ii) receives and updates epoch size values, (iii) identifies epoch boundary packets and sends feedback message to the \inbox up on receiving one. Similar to \inbox's data plane, it can also be implemented using either software or hardware.

\subsection{Prototype}\label{s:impl:prototype}

We now describe our prototype implementation of \name.

\Para{\capinbox data plane} We implement it using Linux \texttt{tc}~\cite{tc}.
We patch the TBF queueing discipline (qdisc)~\cite{tbf} to detect epoch boundary packets and to report them to the control plane using a netlink socket. 
We use the FNV hash function~\cite{fnv-hash}, a non-cryptographic fast hash function with a low collision rate, to compute the packet header hash for identifying epoch boundaries.
This hash function, comprising 4 integer multiplications, is the only additional per-packet work the data plane must perform to support \name; in our experiments, it had negligible CPU overhead.

We patch TBF's \texttt{inner\_qdisc} to support any qdisc-based traffic controller.
\cut{SFQ~\cite{sfq}, FQ-CoDel~\cite{fq-codel} and strict prioritization, in addition to the default FIFO. }
By default, TBF instantaneously re-fills the token bucket when the rate is updated; we disable this feature to avoid rate fluctuations caused by our frequent rate updates. 
Our patch to the TBF qdisc comprises $112$ lines of C.

\Para{\capinbox control plane} We implement it to run in user-space in $1365$ lines of Rust.
We use CCP~\cite{ccp} to run different congestion control algorithms (described next). 
CCP is a platform for expressing congestion control algorithms in an asynchronous format, which makes it a natural choice for our epoch-based measurement architecture. 
The control plane uses \texttt{libccp}~\cite{ccp} to interface with the congestion control algorithm, and  \texttt{libnl} to communicate with the qdisc.

\Para{Congestion control algorithms} We use existing implementations of congestion control algorithms (namely, Nimbus~\cite{nimbus-arxiv}, Copa~\cite{copa} and BBR~\cite{bbr}) on CCP to compute sending rates at the \inbox.  If the algorithm uses a congestion window, the \inbox computes an effective rate of $\frac{\text{CWND}}{\text{RTT}}$ and sets it at the qdisc. 
We validated that our implementation of these congestion control schemes at the \inbox closely follows their implementation at an endhost.

\cut{
\Para{\capoutbox} We implement it using \texttt{libpcap} in $236$ lines of Rust.
}

\subsection{\name Event Loop}\label{s:impl:loop}
Figure~\ref{fig:bundler} provides an overview of how our \name implementation operates on an already-established bundle. 

(1) In the datapath, packets arrive at the \inbox qdisc.
(2) The qdisc determines whether a packet matches the epoch boundary condition (\S\ref{s:measurement}). 
If so, it sends a netlink message to the control plane process running in user-space, and then forwards the packet normally (note the datapath does not send packets to user-space). 
(3) The \outbox observes the same epoch boundary packet via \texttt{libpcap}.
(4) It sends an out-of-band UDP message to the \inbox that contains the hash of the packet and its current state. 
(5) The \inbox receives the UDP message, and uses it to calculate the epochs and measurements as described 
in \S\ref{s:measurement}. 
(6) Asynchronously, the \inbox control plane invokes the congestion control algorithm every $10$ms~\cite{ccp}
via \texttt{libccp},
(7) The \inbox control plane communicates the rate, if updated, to the qdisc
using \texttt{libnl}. 
Finally (8), if the \inbox changes the desired epoch length based on new measurements, it communicates this to the \outbox, also out-of-band.

\section{Evaluation}\label{s:eval}

Given \name's ability to move the in-network queues to the \inbox (as shown earlier in Figure~\ref{fig:design:shift-bottleneck}), we now explore:
\begin{enumerate}[leftmargin=15pt]
    \item Where do \name's performance benefits come from? We discuss this in the context of improving the flow completion times (FCTs) of \name's component flows. (\S\ref{s:eval:fct})
    \item Do \name's performance benefits hold across different scenarios? (\S\ref{s:robust:cross})
    \item Can \name work with different congestion control algorithms (\S\ref{s:eval:cc})?
    \item Are \name's core ideas still applicable with other design decisions? (\S\ref{s:eval:proxy})
    \item Is \name's heuristic (\S\ref{s:queue-ctl:ecmp}) for detecting imbalanced multipath scenarios robust? (\S\ref{s:eval:ecmp})  
    \item Can \name effectively control the queues on real Internet paths? (\S\ref{s:eval:realworld})
\end{enumerate}

\subsection{Experimental Setup}\label{s:eval:setup}

We use network emulation via mahimahi~\cite{mahimahi} to evaluate our implementation of \name in a controlled setting; we present results on real Internet paths in \S\ref{s:eval:realworld}.
There are three $8$-core Ubuntu 18.04 machines in our emulated setup: (1) runs a sender, (2) runs a \inbox, and (3) runs both a \outbox and a receiver.
We disable both TCP segmentation offload (TSO) and generic receive offload (GRO) as they would change the packet headers in between the \inbox and \outbox, which would cause inconsistent epoch boundary identification between the two boxes.
Nevertheless, throughout our experiments CPU utilization on the machines remained below $10$\%.

Unless otherwise specified, we emulate the following scenario.
A many-threaded client generates requests from a request size CDF drawn from an Internet core router~\cite{caida-dataset} and assigns them to one of $200$ server processes.
The workload is heavy-tailed: 97.6\% of requests are 10KB or shorter, and the largest 0.002\% of requests are between $5$MB and $100$MB.
Each server then sends the requested amount of data to the client and we measure the FCT of each such request. 
The link bandwidth at the mahimahi link is set to 96Mbps, and the RTT is set to 50ms. The requests result in an offered load of 84Mbps. 

The endhost runs Cubic~\cite{cubic}, and the \inbox runs Copa~\cite{copa} (we test other schemes in \S\ref{s:eval:cc}) with Nimbus~\cite{nimbus-arxiv} for cross traffic detection.
The \inbox schedules traffic using the Linux kernel implementation of Stochastic Fairness Queueing (SFQ)~\cite{sfq}, though we briefly evaluate other policies in \Sec{s:eval:fct}.
Each experiment is comprised of 1,000,000 requests sampled from this distribution, across 10 runs each with a different random seed.

\subsection{Understanding Performance Benefits}\label{s:eval:fct}

We first present results for a simplified scenario without any cross-traffic, \ie all traffic traversing through the network is generated by the same customer and is, therefore, part of the same bundle. 
This scenario highlights the benefits of using \name when the congestion on the bottleneck link in the network is self-inflicted. We explore the effects of congestion due to other cross-traffic in \S\ref{s:robust:cross}.

\begin{figure}
    \centering
\begin{knitrout}
\definecolor{shadecolor}{rgb}{0.969, 0.969, 0.969}\color{fgcolor}
\includegraphics[width=\maxwidth]{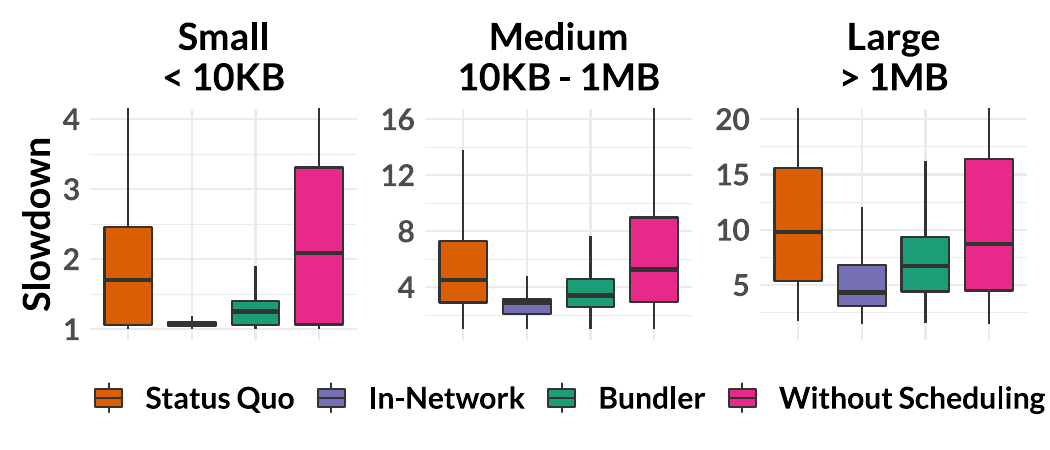} 

\end{knitrout}
    \caption{\name achieves 28\% lower median slowdown. 
The three graphs show FCT distributions for the indicated request sizes: smaller than 10KB, between 10KB and 1MB, and greater than 1MB.  Note the different y-axis scales for each group of request sizes. Whiskers show $1.25 \times$ the inter-quartile range. 
For both \name and In-Network, performance benefits come from preventing short flows from queueing behind long ones.
Thus, \name's aggregate congestion control by itself is not enough; if we configure \name to use FIFO scheduling, the FCTs worsen compared to the status quo.
    }\label{fig:eval:best}
\end{figure}
\newcommand{\overviewBenefitsBaselineMedian}{1.76\xspace}
\newcommand{\overviewBenefitsBaselineTail}{79.37\xspace}
\newcommand{\overviewBenefitsBundlerMedian}{1.26\xspace}
\newcommand{\overviewBenefitsBundlerTail}{41.38\xspace}
\newcommand{\overviewBenefitsOptimalMedian}{1.07\xspace}
\newcommand{\overviewBenefitsOptimalTail}{27.49\xspace}
\newcommand{\overviewBenefitsBundlerMedianImprovement}{28\%\xspace}

\newcommand{\baseline}{Status Quo\xspace}
\newcommand{\optimal}{In-Network\xspace}

\newcommand{\bigexpBundlerElasticSlowdown}{2.6251566\%\xspace}
\newcommand{\bigexpNoBundlerElasticSlowdown}{2.3453956\%\xspace}
\newcommand{\bigexpElasticSlowdownWorseness}{12\%\xspace}

\begin{figure*}
\begin{centering}
\includegraphics[width=\textwidth]{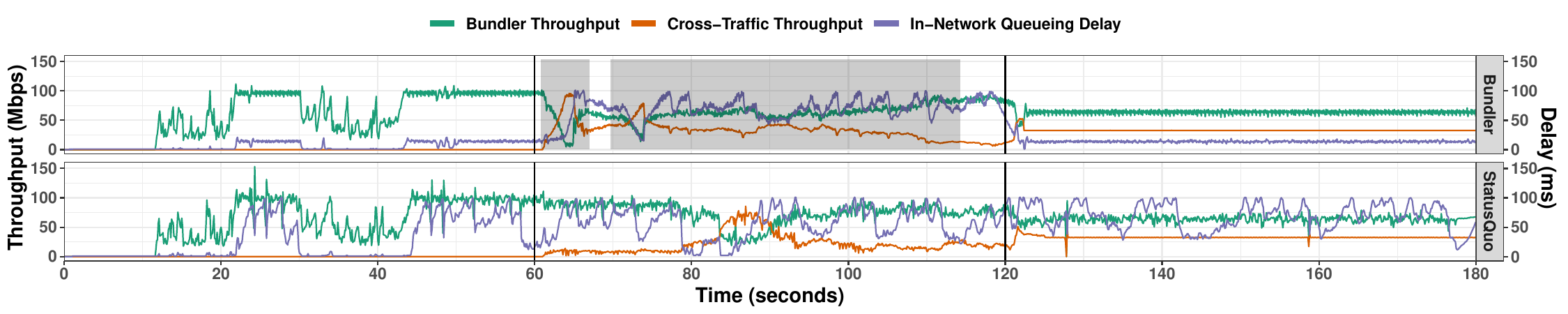}
\end{centering}
\includegraphics[width=0.95\textwidth]{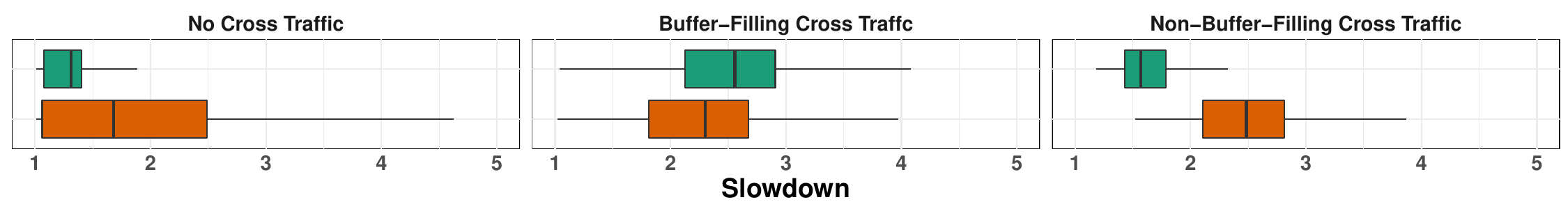}
\caption{\name's scheduling ability depends on the characteristics of the cross traffic over time. In this experiment, there are 3 periods: from 0 to 60 sec., there is no competing traffic, from 60 to 120 sec. there is buffer-filling cross traffic, and from 120 to 180 sec. there is non-buffer-filling cross traffic. The box-plots below each period show the distribution of short flow FCTs during that time. During the period with buffer-filling cross traffic, \name detects its presence and competes fairly. The shaded region indicates time \name spent in buffer-filling cross-traffic mode (\Sec{s:buffer-filling}).}\label{fig:eval:bigexp}
\end{figure*}

\Para{Using \name for fair queueing}
In this section, we evaluate the benefits provided by doing fair queuing at the \name, and use median slowdown as our metric, where the ``slowdown'' of a request is its completion time divided by what its completion time would have been in an unloaded network. A slowdown of $1$ is optimal, and lower numbers represent better performance.

We evaluate three configurations: 
(i) The ``\baseline'' configuration represents the status quo: the \inbox simply forwards packets as it receives them, and the mahimahi bottleneck uses FIFO scheduling.
(ii) The ``\optimal'' configuration deploys fair queueing
at the mahimahi bottleneck.\footnote{
We implement this scheme by modifying mahimahi (our patch comprises $171$ lines of C++) to add a packet-level fair-queueing scheduler to the bottleneck link.}
Recall from \S\ref{s:intro} that this configuration is not deployable.
(iii) The default \name configuration, that uses stochastic fair queueing~\cite{sfq} scheduling policy at the \inbox, and (iv) Using \name with FIFO (without exploiting scheduling opportunity).

Figure~\ref{fig:eval:best} presents our results. 
The median slowdown (across all flow sizes) decreases from \overviewBenefitsBaselineMedian 
for Baseline to \overviewBenefitsBundlerMedian 
with \name, \overviewBenefitsBundlerMedianImprovement
lower. 
\optimal's median slowdown is a further 15\% lower then \name: \overviewBenefitsOptimalMedian.
Meanwhile, in the tail, \name's $99\%$ile slowdown is \overviewBenefitsBundlerTail, which is 48\% lower than the \baseline's \overviewBenefitsBaselineTail. \optimal's $99\%$ile slowdown is \overviewBenefitsOptimalTail.

\paragrapha{Using \name for other policies} We additionally evaluated other scheduling and queue management policies with \name. We omit detailed results for brevity, and present a few highlights. With FQ-CoDel~\cite{fq-codel}, \name can achieve 97\% lower median end-to-end RTTs and 89\% lower 99\%ile RTTs.  By strictly prioritizing one traffic class over another, \name results in 65\% lower median FCTs for the higher-priority class. 

\cut{
\subsection{Applying Different Scheduling Policies}\label{s:eval:policies}

\an{Reduce this subsection to ~one paragraph and roll into 7.2}

In addition to improving FCTs, \name can achieve low packet delay, perform strict prioritization, and rate fairness.

\paragrapha{Achieving Improved Flow Completion Times} \S\ref{s:eval:fct} shows how enabling SFQ at the \name improves the median slowdown by \overviewBenefitsBundlerMedianImprovement.

\paragrapha{Achieving Low Packet Delays}
We enable CoDel~\cite{fq-codel} at the \inbox to lower the packet delays, and test it for a single large backlogged flow using the setup described in \S\ref{s:eval}.
CoDel adds ECN marks to packets in fair-queue buckets which exceed a queue length threshold. 
As a result, endhosts cut their windows earlier, thus reducing their self-inflicted delay within their fair-queue bucket.
We measure the resulting distribution of RTTs seen by the endhost connections with \name and \baseline in Table~\ref{t:eval:codel}.

\newcommand{\delaysImprovement}{97\%\xspace}

\begin{table}[h]
\begin{center}
\begin{tabular}{c|c|c}
Scheme     &  Median RTT                          &    $99$\%ile  RTT                      \\
\hline
Bundler    &  4.25   &    15.32   \\
\baseline  &  122.2  &    151.53
    \label{fig:eval:lowdelays}
\end{tabular}
\end{center}
    \vspace{-10pt}
    \caption{Using CoDel reduces end-to-end packet delays.}\label{t:eval:codel}
\end{table}

As is expected from CoDel, \name in this experiment, achieves \delaysImprovement lower median packet delay than \baseline.

\paragrapha{Strict Prioritization}\label{s:eval:strictprio}
We uniformly divide the web request distribution described in \S\ref{s:eval:setup} into two equally sized classes, one of which is given a higher priority over the other. 
The results are presented in Table~\ref{t:eval:prio}.

\begin{table}[h]
\begin{center}
\begin{tabular}{c|c|c}
Scheme     &  Median Slowdown                           &  $99$\%ile Slowdown                        \\
\hline
Bundler (Prio.)   &  1.07 &  10.52  \\
Bundler (SFQ)     &  \overviewBenefitsBundlerMedian  &  \overviewBenefitsBundlerTail  \\
\baseline  &  3.07  &  201.72
    \label{fig:eval:strict-prio}
\end{tabular}
\end{center}
    \vspace{-10pt}
\caption{Using strict prioritization further reduces FCTs for the higher-priority class of traffic.}\label{t:eval:prio}
\end{table}
\newcommand{\strictPrioTailImprovementOverFq}{47\%\xspace}
\newcommand{\strictPrioImprovement}{65\%\xspace}

Using a priority scheduler (we use the \texttt{pfifo\_fast} qdisc) at \inbox improves the flow completion times for the higher-priority class compared to \baseline.
Furthermore, prioritization achieves \strictPrioTailImprovementOverFq lower $99$\%ile FCT for the higher priority traffic class, when compared to using fair queueing.

\begin{figure}
    \centering
\begin{knitrout}
\definecolor{shadecolor}{rgb}{0.969, 0.969, 0.969}\color{fgcolor}
\includegraphics[width=\maxwidth]{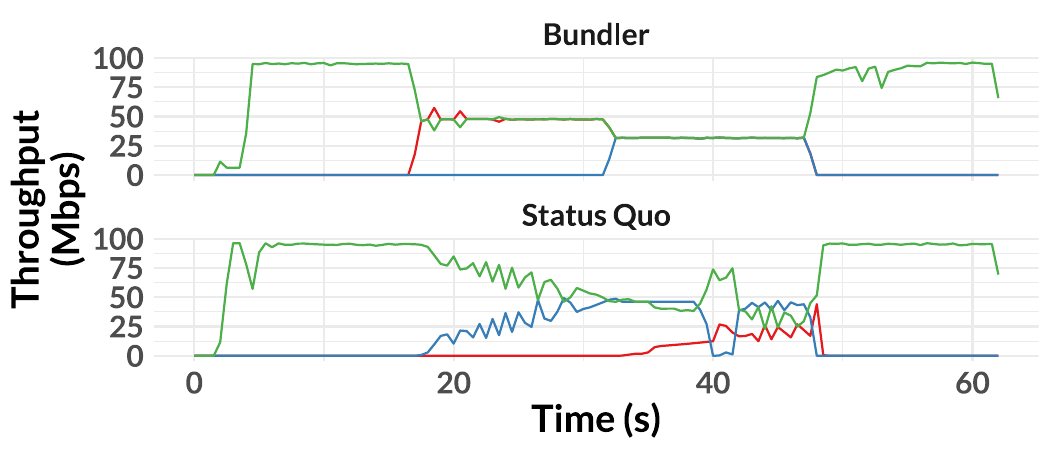} 

\end{knitrout}
    \caption{\name with SFQ achieves fair and stable rates.}
    \label{fig:eval:waterfall}
\end{figure}

\paragrapha{Rate Fairness and Stability}\label{s:eval:waterfall}
We next use our default SFQ scheduler to achieve fairness and rate stability. We start three backlogged flows at different times (0s, 15s, and 30s). Figure~\ref{fig:eval:waterfall} shows that \name converges to fair and stable rates faster than the \baseline.

\begin{figure}
    \label{fig:eval:video}
    \centering
    \begin{subfigure}[b]{0.5\textwidth}
        \includegraphics[width=\textwidth]{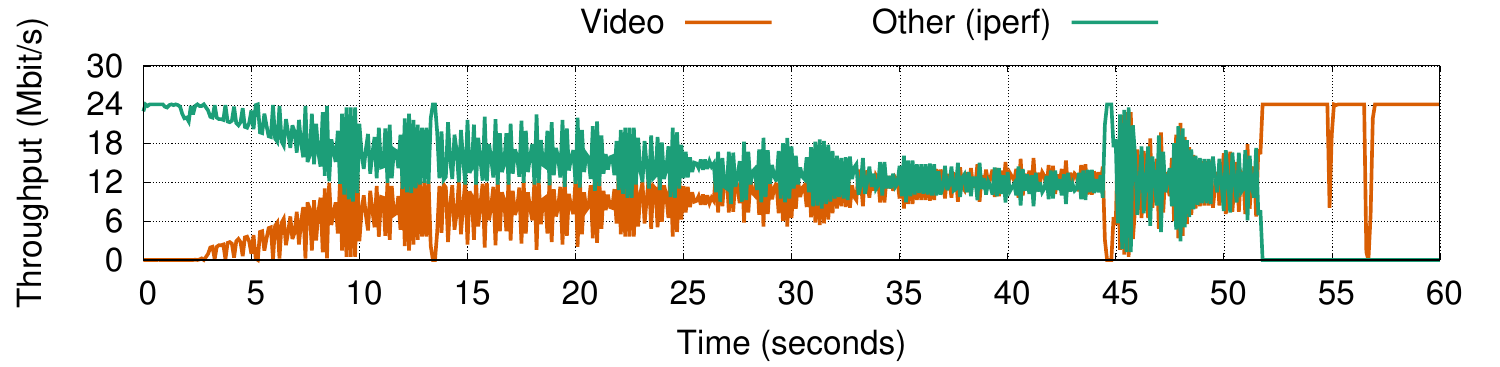}
        \caption{Without \name}\label{fig:eval:video:nobundle}
    \end{subfigure}
    \begin{subfigure}[b]{0.5\textwidth}
        \includegraphics[width=\textwidth]{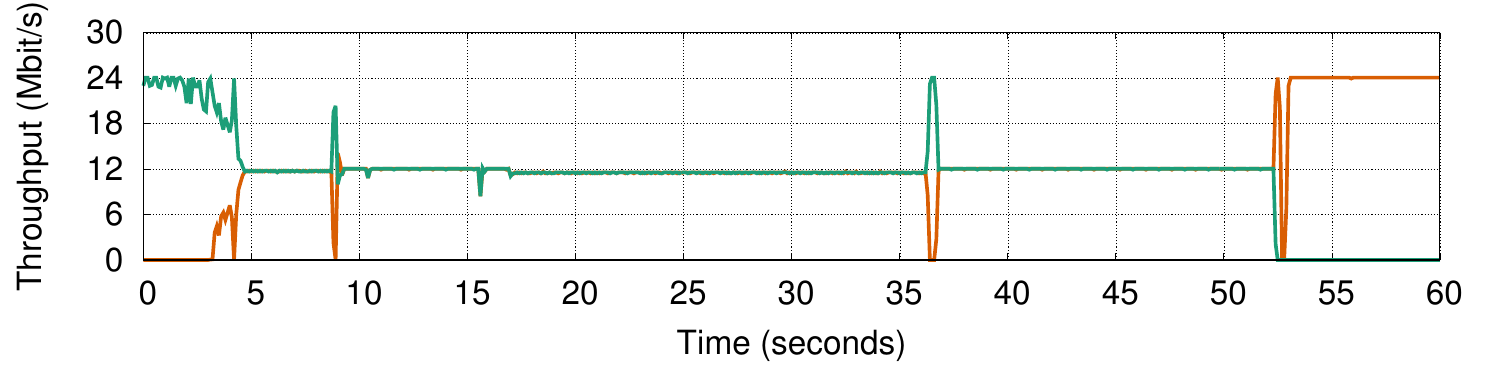}
        \caption{With \name}\label{fig:eval:video:bundle}
    \end{subfigure}
    \caption{Without \name, the video traffic experiences highly variable throughput, which prevents the ABR algorithm from realizing it could sustain a higher bitrate. \name helps the video flow to quickly converge to the fair rate and stay there, which allows the ABR algorithm to choose the maximum sustainable bitrate.}
\end{figure}

\paragrapha{Rate Stability}\label{s:eval:ratestable}
\fc{come back to this}
In Figure~\ref{fig:eval:video}, we run a persistently backlogged flow over a 24Mbps link and then
after 3 seconds start a client attempting to stream a 4k video from a server that supports adaptive
bitrate selection. Without \name (a), the video stream experiences highly variable throughput and 
takes 30 seconds to converge to a fair share of the link. In contrast, with \name (b), the video
stream converges to its fair share within 2 seconds and is able to maintain that rate for the
entirety of the stream. This stability provides the best scenario for the ABR algorithm to select
the highest possible bitrate and thus maximize QoE.
}

\paragrapha{Aggregate congestion control is not enough} It is important to note that \name's congestion control by itself (\ie running FIFO scheduling) is not a means of achieving improved performance. 
To see why this is the case, recall that \name does not modify the endhosts: they continue to run the default Cubic congestion controller, which will probe for bandwidth until it observes loss.
Indeed, the packets endhost Cubic sends beyond those that the link can transmit must queue somewhere in the network or get dropped. 
Without \name, they queue at the bottleneck link;
with \name, they instead queue at the \inbox. 
In addition, the delay-based congestion controller at \inbox also maintains a small standing queue at the bottleneck link (which can be seen in Figure~\ref{fig:design:shift-bottleneck}) to avoid under-utilization, which increases the end-to-end-delays slightly. 
Therefore, doing the FIFO scheduling at the \name, as is done by the \baseline, results in slightly worse performance.


\subsection{Impact of Cross Traffic}\label{s:robust:cross}

Can \name successfully revert to status-quo performance in the presence of buffer-filling cross traffic, then resume providing benefits once that cross traffic leaves?
In Figure~\ref{fig:eval:bigexp}, we show this scenario.
At first, the link is occupied only by \name's traffic, similar to the setup described in \S\ref{s:eval:setup}.
At time $t=60$~sec, a buffer-filling cross traffic flow arrives.
\name detects its presence (indicated by the gray shading) and starts pushing more packets into the network to compete fairly, reverting back to performance that is slightly worse than Status Quo (median FCT for short flows is \bigexpElasticSlowdownWorseness higher). 
Performance is slightly worse because of the $10$ms queue that \name continues to maintain at its \inbox for active probing to detect the cross-traffic's departure, as described in \S\ref{s:queue-ctl}.
\footnote{We believe the benefits provided by \name in the more common regime with no competing buffer-filling cross traffic are substantial enough to make up for slight degradation in these specific scenarios.}
At time $t=120$sec, the buffer-filling flow stops and non-buffer-filling traffic starts, generated from the same distribution as \name as described in \S\ref{s:eval:setup}.
\name correctly detects that it is safe to resume delay-control, and continues providing scheduling benefits.
For the remainder of this subsection, we present three micro-benchmarks which dig deeper into the latter two scenarios, where cross traffic can affect \name's performance. 
We present results with both Nimbus and Copa being used as the congestion control algorithm at the \inbox. 
\cut{In \S\ref{s:eval:offeredload} we present further results on how well \name retains benefits with varying amounts of offered load.}

\begin{figure}
    \centering
\begin{knitrout}
\definecolor{shadecolor}{rgb}{0.969, 0.969, 0.969}\color{fgcolor}
\includegraphics[width=\maxwidth]{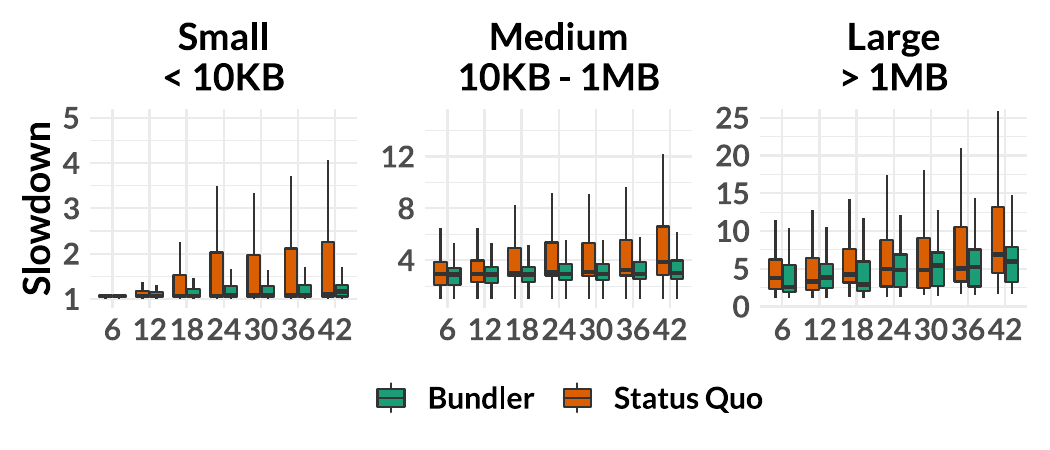} 

\end{knitrout}
    \caption{Against cross traffic comprising of short lived flows. \name offers 48Mbps of load to the bottleneck queue. The cross traffic's offered load increases along the x-axis, while \name{}'s offered load remains fixed.}
    \label{fig:robust:cr-inelastic}
\end{figure}

\paragrapha{Mix of flow sizes} 
We first consider in Figure~\ref{fig:robust:cr-inelastic} the case \name traffic is most likely to encounter, where the cross traffic comprises of finite-length flows up to a few MBs.
We draw both \name's traffic and the cross traffic from the same measured distribution of web requests described in \S\ref{s:eval:setup}.
We fix \name's offered load at a constant $48$ Mbit/s and vary the cross traffic's offered load from $6$ to $42$ Mbit/s.

While flows are often short, they sometimes exit slow start. With sufficient offered load, they can cause queueing in the aggregate. 
Observe that the \baseline FCTs increase steadily as the cross traffic's offered load increases: this is due to the aggregate queueing effect.
When this happens, Bundler's delay-based rate controller could temporarily lower its rate below the aggregate fair share of the bundled traffic.
Importantly, this throughput reduction is short-lived because the queueing is short-lived, and long-term Bundler throughput is not reduced.
We believe that this trade-off (short-term throughput reduction for better delay) is a good one. 
The lower delay helps the short flows in the bundle, while the large flows in the bundle are not affected by the short-term throughput reduction. 
``Mid-sized'' flows in the bundle can be affected if \name sacrifices throughput for too long. 
By design, however, \name detects such cross-traffic and disables its delay-control mechanism in response.

\begin{figure}
    \centering
\begin{knitrout}
\definecolor{shadecolor}{rgb}{0.969, 0.969, 0.969}\color{fgcolor}
\includegraphics[width=\maxwidth]{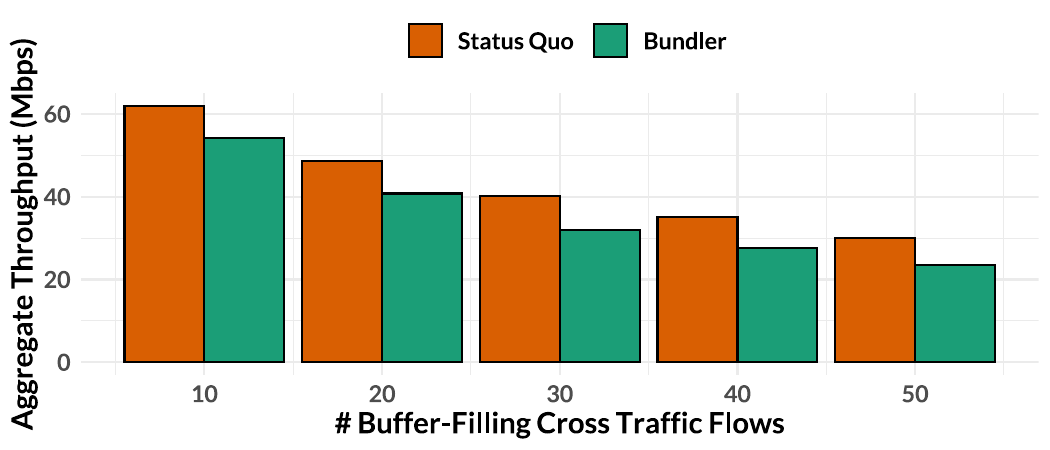} 

\end{knitrout}
    \caption{Varying number of competing buffer-filling cross traffic flows. \name controls a fixed 20 buffer-filling flows in each case.}
    \label{fig:robust:cr-elastic}
\end{figure}

\newcommand{\bundlerElasticTputWorseness}{18\%\xspace}
\newcommand{\bundlerElasticTputWorsenessTen}{12\%\xspace}
\newcommand{\bundlerElasticTputWorsenessFifty}{22\%\xspace}

\paragrapha{Persistent elastic flows} 
We now evaluate how \name's throughput is impacted due to competition from varying amounts of persistent elastic cross-traffic. 
As discussed in \S\ref{s:queue-ctl}, we believe this synthetic scenario is rare in practice, but when it does occur, \name cannot provide benefits, and since it must ``hold back'' some queue to detect when the cross-traffic subsides, its traffic will experience RTT inflation.
Indeed, Figure~\ref{fig:robust:cr-elastic} shows that 
the component flows in the bundle experience  \bundlerElasticTputWorseness less throughput on average. 
The impact varies from \bundlerElasticTputWorsenessTen lower throughput with 10 competing flows to \bundlerElasticTputWorsenessFifty lower with 50. 

\begin{figure}[t]
    \centering
\begin{knitrout}
\definecolor{shadecolor}{rgb}{0.969, 0.969, 0.969}\color{fgcolor}
\includegraphics[width=\maxwidth]{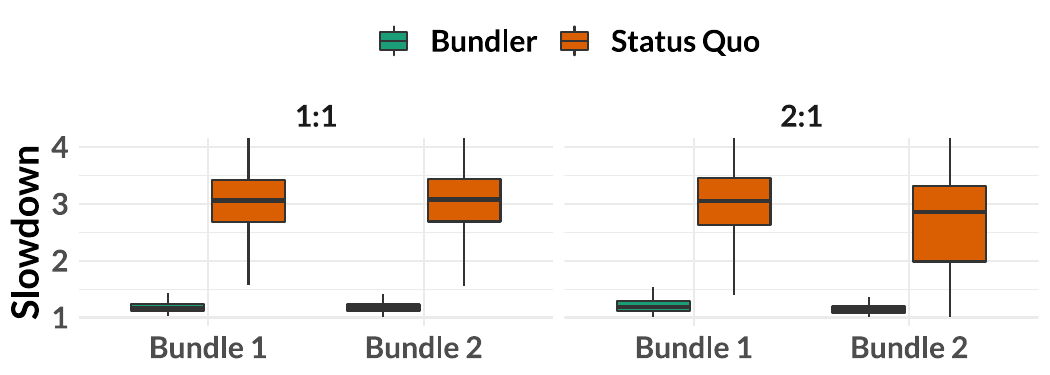} 

\end{knitrout}
    \caption{Competing traffic bundles. In both cases, the aggregate offered load is 84Mbps, as in Figure~\ref{fig:eval:best}. For "1:1", we evenly split the offered load between the two Bundles; for "2:1", one bundle has twice the offered load of the other. In both cases, each bundle observes improved median FCT compared to its performance in the baseline scenario.}
    \label{fig:robust:twobundler}
\end{figure}

\paragrapha{Competing Bundles} Last, we evaluate the case where flows from multiple bundles compete with one another. 
In Figure~\ref{fig:robust:twobundler}, we show the performance with two bundles of traffic competing with one another at the same bottleneck link. 
Both bundles comprise of web requests along with a backlogged Cubic flow. 
Both bundles maintain low queueing in the network and successfully control the queues at the \inbox.
Thus, \name provides benefits for both bundles, even when the amount of traffic in each bundle is different.  

\subsection{Impact of Congestion Control}\label{s:eval:cc}

We now evaluate the impact of a different congestion control algorithm running at the \inbox and at the endhosts.

\begin{figure}[t]
    \centering
\begin{knitrout}
\definecolor{shadecolor}{rgb}{0.969, 0.969, 0.969}\color{fgcolor}
\includegraphics[width=\maxwidth]{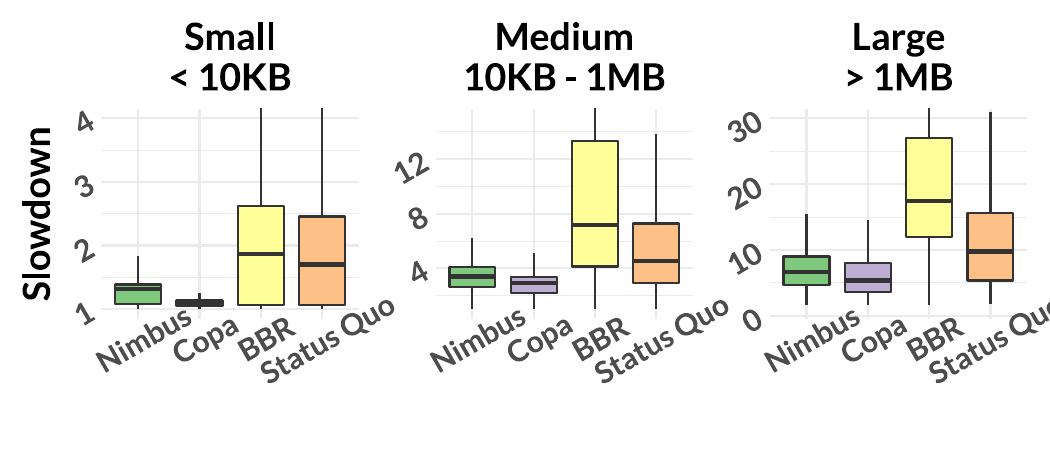} 

\end{knitrout}
    \caption{Choosing a congestion control algorithm at \name remains important, just as it is at the end-host. Note the different y-axis scales for each group of request sizes.}
    \label{fig:eval:cc}
\end{figure}
\newcommand{\ccCopaMedian}{1.09\xspace}
\newcommand{\ccNimbusMedian}{1.32\xspace}
\newcommand{\ccBBRMedian}{1.91\xspace}
\newcommand{\ccBaselineMedian}{1.76\xspace}

\Para{\capinbox congestion control} So far we have evaluated \name by running Copa~\cite{copa} at the \inbox. 
Figure~\ref{fig:eval:cc} shows \name's performance with other congestion control algorithms (namely, Nimbus's BasicDelay~\cite{nimbus-arxiv} and BBR~\cite{bbr}), and using SFQ scheduling. 
We find that using BasicDelay provides similar benefits over \baseline as Copa. 
BBR, on the other hand, performs slightly worse than \baseline. 
This is because it pushes packets into the network more aggressively than the other schemes, resulting in a bigger in-network queue.
This, combined with the queue built at the \name, results in the endhosts experiencing higher queueing delays than \baseline. This shows that the choice of congestion control algorithm, and its ability to maintain small queues in the network, plays an important role. 

\Para{Endhost congestion control} We used Cubic congestion control at the endhosts for our experiments so far. When we configure endhosts to use Reno or BBR, \name's benefits remain: \name achieves 58\% lower FCTs in the median compared to the updated \baseline where the endhosts use BBR. 
This shows that \name is compatible with multiple endhost congestion control algorithms.

\cut{
\begin{figure}
    \centering
\begin{knitrout}
\definecolor{shadecolor}{rgb}{0.969, 0.969, 0.969}\color{fgcolor}
\includegraphics[width=\maxwidth]{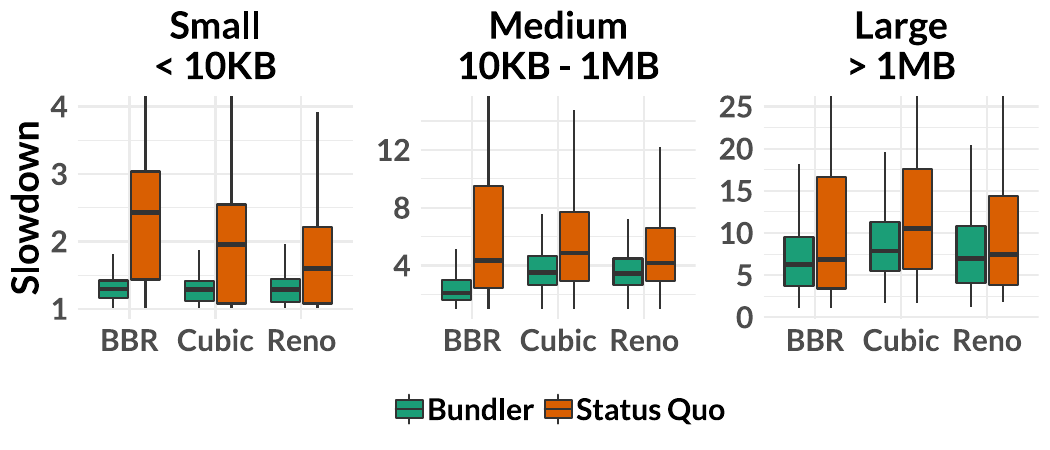} 

\end{knitrout}
    \caption{\name still provides benefits when the endhosts use different congestion control algorithms.}
    \label{fig:eval:traffic}
\end{figure}

\Para{Endhost congestion control} 
We used Cubic congestion control at the endhosts for our experiments so far. When we configure endhosts to use Reno or BBR (as implemented in Linux $4.13$), \name's benefits remain (Figure~\ref{fig:eval:traffic}).
This shows that \name is compatible with multiple endhost congestion control algorithms.
}

\cut{
\begin{Appendix}
\section{Varying Offered Load}\label{s:eval:offeredload}
Naturally, if a link is less congested, scheduling the packets that traverse it will have less benefit. Accordingly, as the offered load is reduced, we would expect the gains from scheduling to diminish. 
We now use the web request distribution described in \S\ref{s:eval:setup} to generate a load of 50\% ($48$Mbps), 75\% ($72$Mbps) and 87.5\% ($84$Mbps) of the bottleneck link bandwidth. Our results in Figure~\ref{fig:eval:offeredload} show that as the offered load is decreased, the benefits of \name reduce. This is because if a link is less congested, scheduling the packets that traverse it will have less benefit.
\end{Appendix}
}

\subsection{Terminating TCP Connections}\label{s:eval:proxy}

Although our \name prototype does not terminate connections (as discussed in \S\ref{s:design:whichcc}), we note that terminating connections does provide one key advantage: the end-to-end congestion controller will observe a smaller RTT, since the proxy can acknowledge its segments much faster than the original receiver. 
This enables rapid window growth at the endhosts.
While there are, of course, operational concerns with managing the resulting queue, it does provide additional scheduling opportunities as well as faster ramp-up for midsized connections.

\begin{figure}
    \centering
\begin{knitrout}
\definecolor{shadecolor}{rgb}{0.969, 0.969, 0.969}\color{fgcolor}
\includegraphics[width=\maxwidth]{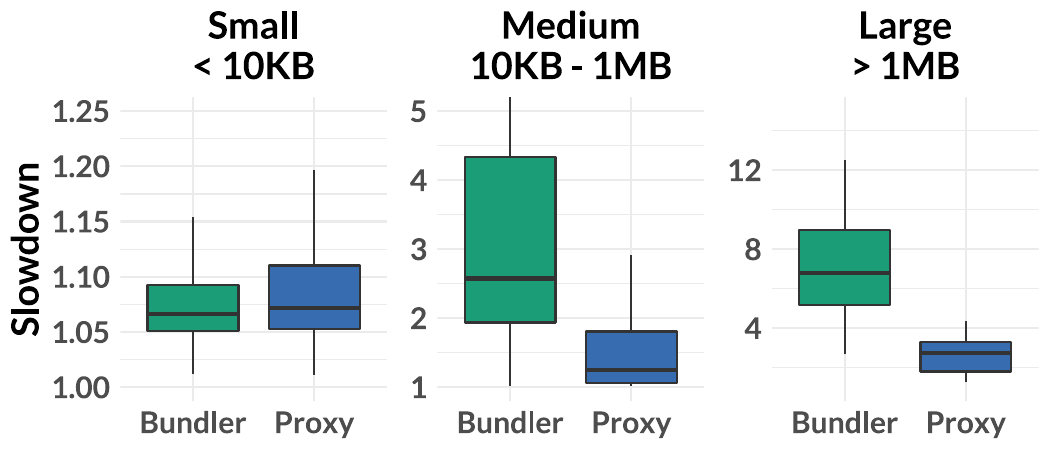} 

\end{knitrout}
    \caption{A proxy-based implementation of \name could yield further benefits to the long flows. Note the different y-axis scales for each group of request sizes.}
    \label{fig:eval:proxy}
\end{figure}

How much benefit, then, could a proxy-based \name provide?
To evaluate this, we emulate an idealized TCP proxy by modifying the endhosts to maintain a constant congestion window of $450$ packets---slightly larger than the bandwidth-delay product in our setup---and increasing the buffering at the \inbox to hold these packets. 
The other aspects of \name remain unchanged.
The result is in Figure~\ref{fig:eval:proxy}. 

For the short requests which never leave TCP slow start, terminating TCP connections does not yield additional benefits: with or without termination, they finish in a few RTTs.
For medium-to-long requests, terminating TCP connections yields additional benefits since they no longer incur the penalty of window growth.
Therefore, a site may benefit from proxying TCP connections at \name if its traffic pattern contains many medium-sized flows which benefit from fast ramp-up.

\subsection{Multipath Detection}\label{s:eval:ecmp}

As described in \S\ref{s:queue-ctl:ecmp}, when the ratio of out-of-order to in-order measurements is above a certain threshold, it indicates that \name's component flows are likely traversing multiple imbalanced paths. To evaluate the extent to which this heuristic corresponds with imbalance, we re-run the emulation experiment from Figure~\ref{fig:eval:bigexp} for a variety of network conditions (bottleneck bandwidth ranging from 12 to 96 Mbps, end-to-end RTTs ranging from 10 to 300 ms, and bottleneck load-balancing from 1 to 32 paths) and consider the average value reported by the heuristic over each experiment. The maximum value reported across all experiments with a single path was 0.4\%, while the minimum value reported across all experiments with 2-32 paths was 20\%, two orders of magnitude greater. Thus, this heuristic provides a very clear separation between single and multiple path scenarios and a simple threshold is sufficient. 



\section{Real Internet Paths}\label{s:eval:realworld}

\begin{figure}
    \centering
\begin{knitrout}
\definecolor{shadecolor}{rgb}{0.969, 0.969, 0.969}\color{fgcolor}
\includegraphics[width=\maxwidth]{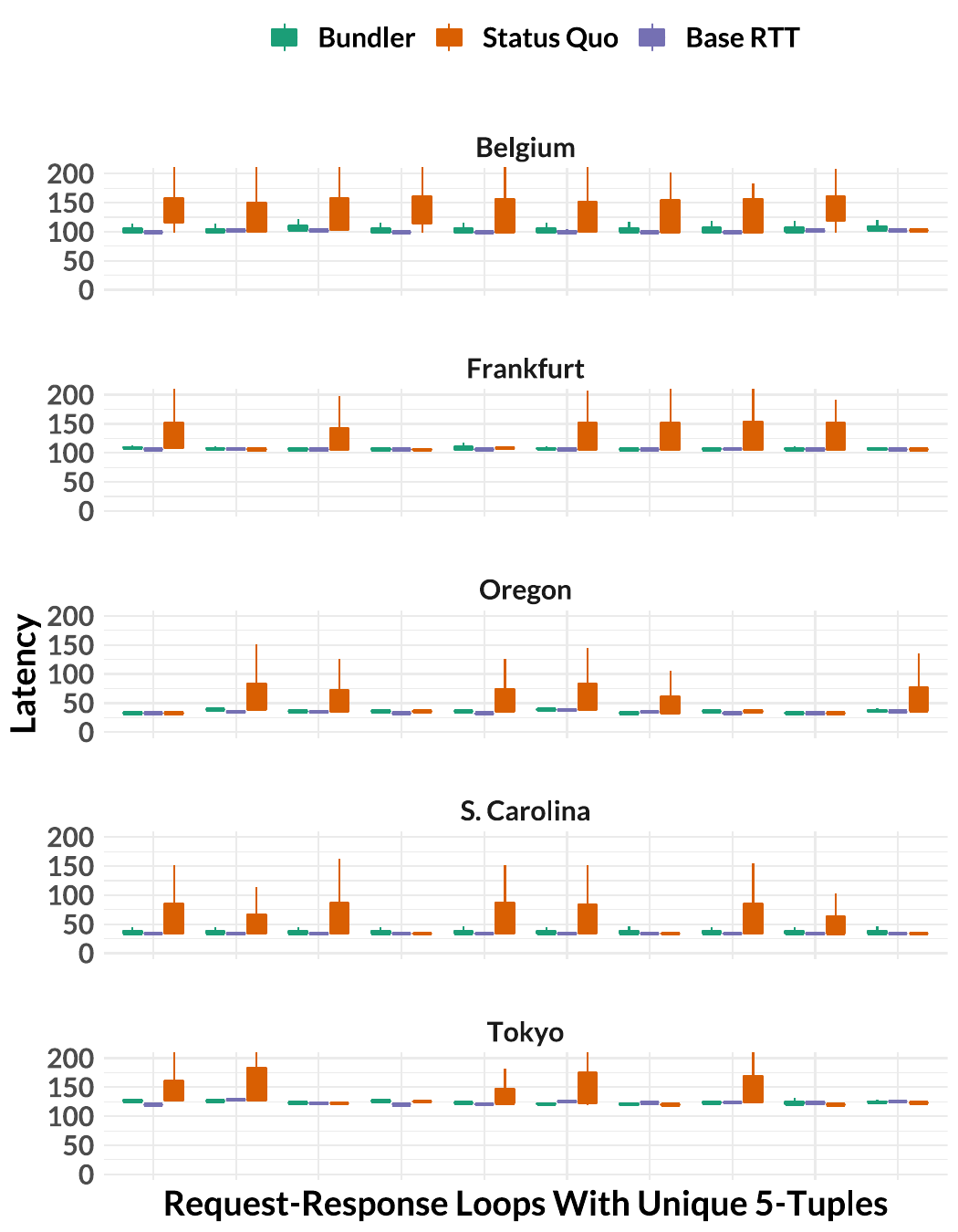} 

\end{knitrout}
\newcommand{\realworldMedianLatencyImprovement}{57\%\xspace}
\newcommand{\realworldAvgBwRatio}{1\%\xspace}
\caption{On 5 real-Internet paths, \name achieves lower latencies than Status Quo for latency-sensitive traffic. Each bar depicts an individual 5-tuple; load-balancing in the Internet prevents queueing for some 5-tuples. \name still offers scheduling for paths with queueing (achieving \realworldMedianLatencyImprovement lower latencies overall) while achieving overall throughput within \realworldAvgBwRatio of that in the Status Quo scenario.}
    \label{fig:eval:realworld}
\end{figure}

\newcommand{\realworldMedianLatencyImprovement}{57\%\xspace}
\newcommand{\realworldAvgBwRatio}{1\%\xspace}

We next evaluate our prototype implementation on real Internet paths to demonstrate that \name can effectively shift queues in practical settings.

\Para{Experiment Setup} 
We deploy \name (\inbox) in a GCP datacenter in Iowa and generate traffic from multiple different machines in this datacenter (as detailed below). 
The generated traffic is sent to multiple machines in five different GCP datacenters (in Belgium, Frankfurt, Oregon, South Carolina, and Tokyo). We configured GCP to route traffic over the public Internet rather than a private network. 
We deploy a \name (\outbox) in each of these receiving datacenters, thus resulting in a total of five bundles spanning different regions of the globe.

We evaluate two different workloads in this setup: (i) Each bundle comprising of 10 parallel closed-loop 40 bytes UDP requests, where the sender issues a new request every time it receives a response. We measure the request-response RTTs in this workload to use as a baseline (and call them Base RTTs). (ii) We add 20 backlogged (\texttt{iperf}) flows to the above workload in each bundle. We run this workload both with and without \name and measure the UDP request-response RTTs (represented as \name and \baseline respectively). Effective SFQ across all flows with \name should not inflate the base request-response RTT.
We verified that the backlogged senders achieve similar throughput in all cases (2-4Gbit/s on these paths) both with and without \name, and that the \name machine in Iowa is not a bottleneck itself. 

\Para{Result} 
Figure~\ref{fig:eval:realworld} shows, for each of the five bundles, the resulting RTT distributions for each of the ten request-response loops (with the 5 tuples in UDP/IP headers differing across all ten). 
We make two key observations: (i) The \baseline RTTs are significantly higher than the Base RTT, which indicates significant queueing outside of either site's control. (ii) \name is able to move these queues and enforce SFQ scheduling effectively, resulting in request-response RTTs comparable to Base RTTs, and \realworldMedianLatencyImprovement smaller than \baseline at the median.

\Para{Explanation}
Observation (i) above indicates that all of the conditions in \S\ref{s:deploy} held during our experiment and that queues were indeed building outside of our control. 
One possibility is that these queues built up at an egress rate limiter imposed by GCP. 
However, our throughput measurements suggest that this is unlikely\footnote{We lack visibility into Google's network and thus were unable to determine the true location of the bottleneck in this experiment.}; we used \texttt{n1-standard-2} machines, which have a maximum possible egress of 10Gbps each, but our backlogged senders achieved only 2-4Gbps. 
Nevertheless, even if queues did form at GCP's rate limiter, this represents a scenario where \name is useful: 
an operator deploying an application between multiple cloud regions could use \name to enforce scheduling policies on their traffic without negotiating their policy with the cloud provider or knowing where the bottleneck occurs. 

\section{Discussion}\label{s:discussion}

\paragrapha{Composability} Bundles are naturally \emph{composable}: a sub-site within site A can deploy its own \name to take control of its fraction of the in-network queues, with the site A's \name enforcing a scheduling policy across the bundled traffic from each sub-site.  
For example, a department within an institute may bundle its traffic to a collaborating department in another institute, with the parent institutes bundling the aggregate traffic across multiple departments.

\paragrapha{Scheduling across different bundles at a \inbox} We evaluate benefits of scheduling \emph{within} a bundle. In practice, a given \inbox will see traffic from multiple bundles. Extending different scheduling policies to multiple such bundles can be done trivially.

\paragrapha{Rate allocation across different competing bundles} When multiple bundles (belonging to different sites) compete at the same bottleneck, \name's congestion control would ensure a fair rate allocation across each of these bundles, irrespective of the amount of traffic in them. It, therefore, provides fairness on per-site basis, as opposed to a per-flow basis, making it more robust to popular end-host strategies such as opening multiple connections to increase bandwidth share. 

\section{Conclusion}\label{s:concl}
We have described \name, a new type of middlebox which uses a novel ``inner''  congestion control loop for traffic bundles between two sites to shift the queues from the middle of the network, where it is difficult to unilaterally express traffic control policy, to the site itself, where doing so is tractable. 
\name neither maintains any per-flow state, nor makes any modifications to the packets. 
We demonstrate, using both emulated network experiments and real Internet paths, that it is possible to shift queues and schedule packets to an extent sufficient to enforce well-known scheduling disciplines. 

\cut{In our experiments, we used fair queueing to improve median FCTs by \overviewBenefitsBundlerMedianImprovement, strict prioritization to improve $99$\%ile FCTs of a high-priority traffic class by \strictPrioImprovement, and AQM to reduce end-to-end packet delays by \delaysImprovement.}

\begin{acks}
We thank Srinivas Narayana, Ahmed Saeed, Rachee Singh, the anonymous EuroSys reviewers, and our shepherd Andreas Haeberlen for their helpful discussions and feedback.
This work is supported in part by DARPA contract HR001117C0048 and NSF grants 1526791, 1563826, 2006346, and 1407470.
\end{acks}

\label{p:end}
\end{sloppypar}
\clearpage
\bibliographystyle{abbrv}
\bibliography{ref}
\end{document}